\documentclass[twocolumn]{aastex62}

\usepackage{natbib}
\usepackage{multirow}
\usepackage{amsmath}
\usepackage{xcolor}

\def\rhohi{$\rho_{\rm HI}$}
\def\HI{\ion{H}{1}}

\graphicspath{{./}{figures/}}

\shortauthors{Walter et al.}

\begin{document}

\title{The Evolution of the Baryons Associated with Galaxies Averaged over Cosmic Time and Space}

\correspondingauthor{Fabian Walter}
\email{walter@mpia.de}

\author[0000-0003-4793-7880]{Fabian Walter}
\affil{Max Planck Institute for Astronomy, K\"onigstuhl 17, 69117 Heidelberg, Germany}
\affil{National Radio Astronomy Observatory, Pete V. Domenici Array Science Center, P.O. Box O, Socorro, NM 87801, USA}

\author[0000-0001-6647-3861]{Chris Carilli}
\affil{National Radio Astronomy Observatory, Pete V. Domenici Array Science Center, P.O. Box O, Socorro, NM 87801, USA}

\author[0000-0002-9838-8191]{Marcel Neeleman}
\affil{Max Planck Institute for Astronomy, K\"onigstuhl 17, 69117 Heidelberg, Germany}

\author[0000-0002-2662-8803]{Roberto Decarli}
\affiliation{INAF—Osservatorio di Astrofisica e Scienza dello Spazio, via Gobetti 93/3, I-40129, Bologna, Italy}

\author[0000-0003-1151-4659]{Gerg\"{o} Popping}
\affil{European Southern Observatory, Karl-Schwarzschild-Strasse 2, 85748, Garching, Germany}

\author{Rachel S. Somerville}
\affil{Rutgers University, 136 Frelinghuysen Road, Piscataway, NJ 08854-8019, USA}
\affil{Center for Computational Astrophysics, Flatiron Institute, 162 5th Avenue, New York, NY 10010, USA}

\author[0000-0002-6290-3198]{Manuel Aravena}
\affil{N\'{u}cleo de Astronom\'{\i}a, Facultad de Ingenier\'{\i}a y Ciencias, Universidad Diego Portales, Av. Ej\'{e}rcito 441, Santiago, Chile}

\author[0000-0002-1707-1775]{Frank Bertoldi}
\affil{Argelander-Institut f\"ur Astronomie, Universit\"at Bonn, Auf dem H\"ugel 71, 53121 Bonn, Germany}

\author[0000-0002-3952-8588]{Leindert Boogaard}
\affil{Leiden Observatory, Leiden University, P.O. Box 9513, NL-2300 RA Leiden, The Netherlands}

\author[0000-0003-2027-8221]{Pierre Cox}
\affil{Institut d'Astrophysique de Paris, Sorbonne Universit\'e, CNRS, UMR 7095, 98 bis Blvd. Arago, 75014 Paris, France}

\author{Elisabete da Cunha}
\affil{International Centre for Radio Astronomy Research, The University of Western Australia, 35 Stirling Highway, Crawley WA 6009, Australia}

\author[0000-0002-6777-6490]{Benjamin Magnelli}
\affil{Argelander-Institut f\"ur Astronomie, Universit\"at Bonn, Auf dem H\"ugel 71, 53121 Bonn, Germany}

\author{Danail Obreschkow}
\affil{International Centre for Radio Astronomy Research, The University of Western Australia, 35 Stirling Highway, Crawley WA 6009, Australia}

\author[0000-0001-9585-1462]{Dominik Riechers}
\affil{Cornell University, 220 Space Sciences Building, Ithaca, NY 14853, USA}

\author[0000-0003-4996-9069]{Hans--Walter Rix}
\affil{Max Planck Institute for Astronomy, K\"onigstuhl 17, 69117 Heidelberg, Germany}

\author{Ian Smail}
\affil{Centre for Extragalactic Astronomy, Durham University,  Department of Physics,  South Road, Durham DH1 3LE,  UK}

\author[0000-0003-4678-3939]{Axel Weiss}
\affil{Max-Planck-Institut f\"ur Radioastronomie, Auf dem H\"ugel 69, 53121 Bonn, Germany}

\author[0000-0002-9508-3667]{Roberto J. Assef}
\affil{N\'{u}cleo de Astronom\'{\i}a, Facultad de Ingenier\'{\i}a y Ciencias, Universidad Diego Portales, Av. Ej\'{e}rcito 441, Santiago, Chile}

\author[0000-0002-8686-8737]{Franz Bauer}
\affil{Instituto de Astrofísica, Facultad de Física, Pontificia Universidad Católica de Chile Av. Vicuña Mackenna 4860, 782-0436 Macul, Santiago, Chile}
\affil{Millennium Institute of Astrophysics (MAS), Nuncio Monseñor Sótero Sanz 100, Providencia, Santiago, Chile}
\affil{Space Science Institute, 4750 Walnut Street, Suite 205, Boulder, CO 80301, USA}

\author{Rychard Bouwens}
\affil{Leiden Observatory, Leiden University, P.O. Box 9513, NL-2300 RA Leiden, The Netherlands}

\author{Thierry Contini}
\affil{Institut de Recherche en Astrophysique et Plan\`etologie (IRAP), Universit\'e de Toulouse, CNRS, UPS, F--31400 Toulouse, France}

\author{Paulo C. Cortes}
\affil{Joint ALMA Office, Alonso de Cordova 3107, Vitacura, Santiago, Chile}
\affil{National Radio Astronomy Observatory, Charlottesville, VA 22903, USA}

\author[0000-0002-3331-9590]{Emanuele Daddi}
\affil{Laboratoire AIM, CEA/DSM-CNRS-Universite Paris Diderot, Irfu/Service d’Astrophysique, CEA Saclay, Orme des Merisiers, F-91191 Gif-sur-Yvette cedex, France}

\author[0000-0003-0699-6083]{Tanio Diaz-Santos}
\affil{N\'{u}cleo de Astronom\'{\i}a, Facultad de Ingenier\'{\i}a y Ciencias, Universidad Diego Portales, Av. Ej\'{e}rcito 441, Santiago, Chile}
\affil{Chinese Academy of Sciences South America Center for Astronomy (CASSACA), National Astronomical Observatories, CAS, Beijing 100101, China}
\affil{Institute of Astrophysics, Foundation for Research and Technology-Hellas (FORTH), Heraklion, GR-70013, Greece}

\author[0000-0003-3926-1411]{Jorge Gonz\'alez-L\'opez}
\affil{N\'{u}cleo de Astronom\'{\i}a, Facultad de Ingenier\'{\i}a y Ciencias, Universidad Diego Portales, Av. Ej\'{e}rcito 441, Santiago, Chile}

\author[0000-0002-7054-4332]{Joseph Hennawi}
\affil{Department of Physics, Broida Hall, University of California, Santa Barbara, CA 93106-9530, USA}

\author[0000-0001-6586-8845]{Jacqueline A. Hodge}
\affil{Leiden Observatory, Leiden University, P.O. Box 9513, 2300 RA Leiden, the Netherlands}

\author{Hanae Inami}
\affil{Hiroshima Astrophysical Science Center, Hiroshima University, 1-3-1 Kagamiyama, Higashi-Hiroshima, Hiroshima, 739-8526, Japan}

\author[0000-0001-5118-1313]{Rob Ivison}
\affil{European Southern Observatory, Karl--Schwarzschild--Strasse 2, 85748, Garching, Germany}

\author[0000-0001-5851-6649]{Pascal Oesch}
\affil{Department of Astronomy, University of Geneva, Ch. des Maillettes 51, 1290 Versoix, Switzerland}
\affil{International Associate, Cosmic Dawn Center (DAWN) at the Niels Bohr Institute, University of Copenhagen and DTU-Space, Technical University of Denmark, Copenhagen, Denmark}

\author[0000-0003-1033-9684]{Mark Sargent}
\affil{Astronomy Centre, Department of Physics and Astronomy, University of Sussex, Brighton, BN1 9QH, UK}

\author{Paul van der Werf}
\affil{Leiden Observatory, Leiden University, P.O. Box 9513, NL--2300 RA Leiden, The Netherlands}

\author{Jeff Wagg}
\affil{SKA Organization, Lower Withington Macclesfield, Cheshire SK11 9DL, UK}

\author{L. Y. Aaron Yung}
\affil{Rutgers University, 136 Frelinghuysen Road, Piscataway, NJ 08854-8019, USA}




\vspace{1cm}

\begin{abstract}
We combine the recent determination of the evolution of the cosmic density of molecular gas (H$_2$) using deep, volumetric surveys, with previous estimates of the cosmic density of stellar mass, star formation rate and atomic gas (\ion{H}{1}), to constrain the evolution of baryons associated with galaxies averaged over cosmic time and space. The cosmic \ion{H}{1} and H$_2$ densities are roughly equal at $z\,\sim\,1.5$. The H$_2$ density then decreases by a factor 6$^{+3}_{-2}$ to today's value, whereas the \ion{H}{1} density stays approximately constant. The stellar mass density is increasing continuously with time and surpasses that of the total gas density (\ion{H}{1} and H$_2$) at redshift $z\,\sim\,1.5$. The growth in stellar mass cannot be accounted for by the decrease in cosmic H$_2$ density, necessitating significant accretion of additional gas onto galaxies. With the new H$_2$ constraints, we postulate and put observational constraints on a two step gas accretion process: (i) a net infall of ionized gas from the intergalactic/circumgalactic medium to refuel the extended \ion{H}{1} reservoirs, and (ii) a net inflow of \ion{H}{1} and subsequent conversion to H$_2$ in the galaxy centers. Both the infall and inflow rate densities have decreased by almost an order of magnitude since $z\,\sim\,2$. Assuming that the current trends continue, the cosmic molecular gas density will further decrease by about a factor of two over the next 5\,Gyr, the stellar mass will increase by approximately 10\%, and cosmic star formation activity will decline steadily toward zero, as the gas infall and accretion shut down.
\end{abstract}



\section{Introduction}

The principal goal in galaxy evolution studies is to understand how the cosmic structure and galaxies that we see today emerged from the initial conditions imprinted on the Cosmic Microwave Background (CMB). In the hierarchical structure formation paradigm, galaxies grow both through the smooth accretion of dark matter and baryons, and through distinct mergers of dark matter halos (and their associated  baryons). The accretion of gas eventually leads to the formation of stars in galaxies in the centers of the individual dark matter halos \citep[e.g.][]{white78,blumenthal84,white91}. The winds, UV photons and supernovae from the ensuing star formation, along with possible episodic accretion onto the supermassive black hole at the center (active galactic nuclei), provide effective `feedback' to the surrounding gas. This may -- at least temporarily -- suppress the formation of further stars, or may even expel the cold gas from the centers of the potential wells \citep[e.g.,][]{dekel86,silk98,croton06,somerville08}. Together, this leads to a baryon cycle through different gas phases and galactocentric radii \citep[e.g.,][]{tumlinson17}. Of particular interest in this baryon cycle is the question: how much gas was present both within and around galaxies to explain the formation of stars in galaxies through cosmic times?

Over the past decades, deep sky surveys of star formation and stars in the optical and (near--)infrared bands have put tight constraints on the build--up of the stellar mass in galaxies from early cosmic times to the present \cite[e.g., review by][]{madau14}. In parallel, the atomic hydrogen content has been derived through \ion{H}{1} emission in the local universe \citep[e.g., ][]{zwaan05}, and quasar absorption spectroscopy at high redshift  \cite[e.g., ][]{prochaska09}. The molecular gas content of galaxies, the immediate fuel for star formation, has now also been constrained as a function of redshift through measurements of the molecular transitions of carbon monoxide, CO, as well as the far--infrared dust continuum \cite[e.g., reviews by][]{carilliwalter13, tacconi20, peroux20, hodge20}. These include recent measurements from the ALMA Spectroscopic Survey in the Hubble Ultra--deep Field \citep[ASPECS;][]{decarli19,decarli20,magnelli20}. Together, the available data have now reached the point that we can account for the total cold gas content (\ion{H}{1} and H$_2$) that is associated with galaxies as a function of cosmic time.

In this paper we discuss how these new molecular gas constraints impact our view of the cosmic baryon cycle of galaxies, and, in particular, how they affect our view of  gas accretion to sustain the observed star formation rate density in the centers of galaxies. Throughout this paper we only consider densities that are averaged over cosmic space and wide time bins to characterize the cosmic baryon cycle (Sec.~\ref{baryons}). We argue that such an approach is justified as molecular gas, star formation, and stellar mass are found to be approximately co--spatial in galaxies, and because the averaging times are significantly longer than the physical processes under consideration (Sec.~\ref{schematic}). We thus stress that many conclusions of this paper, including the accretion and inflow rates, will not be applicable to individual galaxies, but only to volume--averaged galaxy samples (Sec.~\ref{discussion}). Given the available observational constraints, we here focus on redshifts below $z\,\sim\,4$ (when the Universe was older than 1.5\,Gyr). 

We adopt a `cosmic concordance cosmology' with the following parameters: a reduced Hubble constant $h\,=$H$_0/(100\,{\rm km s}^{-1} \,{\rm Mpc}^{-1})\,=\,0.7$, a matter density parameter $\Omega_m$\,=\,0.31 (which is the sum of the dark matter density parameter $\Omega_c$\,=\,0.259 and the baryon density parameter $\Omega_b$\,=\,0.048), and a dark energy density parameter $\Omega_\Lambda\,=\,(1-\Omega_m)$\,=\,0.69, similar to Planck constraints \citep{planck15}, and those used in the review on cosmic star formation rates and associated stellar mass build--up by  \citet{madau14}. All the volume--averaged, cosmological densities quoted in this paper are in co--moving units.

\section{A simple schematic}
\label{schematic}
\begin{figure}
\hspace*{-0.7cm}
\centering 
\includegraphics[angle=0,width=1.15\columnwidth]{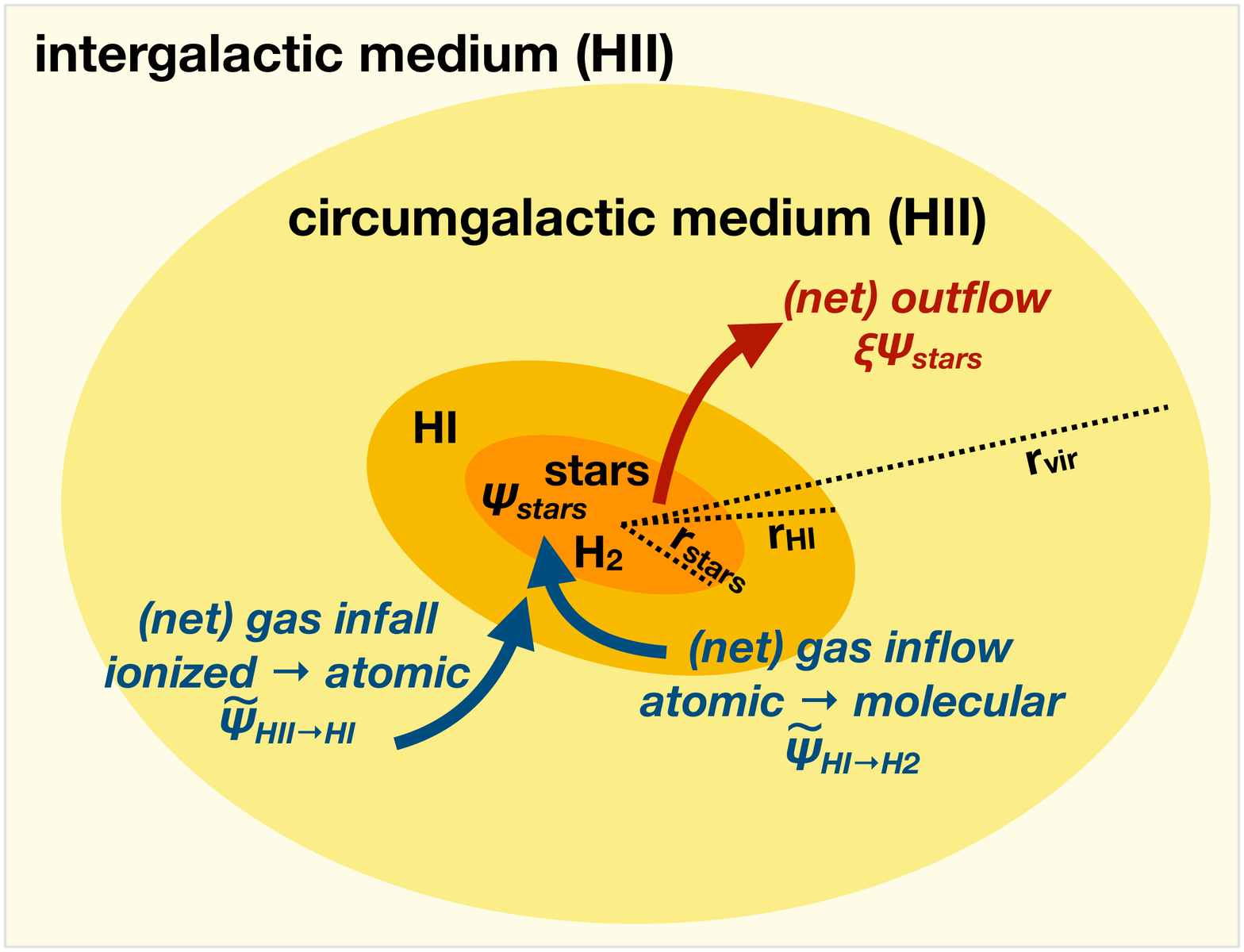}
\caption{Schematic of the different baryonic components that are present within the dark matter halo of a galaxy (defined as $r\,<\,r_{\rm vir}$). The central `disk' region ($r<r_{\rm stars}$), contains the vast majority of stars and molecular gas, and stars form here at a rate $\psi_{\rm stars}$. This region is surrounded by a reservoir of atomic gas (\ion{H}{1}), with $r\,<\,r_{\rm HI}$. The predominantly ionized material (\ion{H}{2}) beyond this radius,  but within r$_{\rm vir}$, constitutes the circumgalactic medium (CGM). Beyond r$_{\rm vir}$ is the realm of the intergalactic medium (IGM). Blue arrows indicate the (net) infall of ionized gas to the \ion{H}{1} reservoir ($\widetilde\psi_{\rm HII\rightarrow HI}$) as well as the (net) inflow of atomic gas to the molecular gas (H$_2$) reservoir ($\widetilde\psi_{\rm HI\rightarrow H2}$). The red arrow indicates the material entrained in outflows that can reach the CGM and possibly the IGM (here assumed to be proportional to $\psi_{\rm stars}$).
\label{fig:schematic}
}
\end{figure}

Fig.~\ref{fig:schematic} shows a schematic of the different baryonic components that are present within the dark matter halo of a galaxy. The central region of the galaxy contains the majority of the stars, molecular gas, and star formation at any given time (Secs.~\ref{stars},~\ref{H2}). In this region, stars form out of giant molecular clouds with a typical timescale of order $10^7$\,yr \citep[e.g.,][]{kawamura09,meidt15,schinner19} and molecular gas is expected to form out of atomic gas on a similar timescale \citep[depending on metallicity, e.g.,][]{fukui09,glover11,clark12,walch15}. These periods are significantly shorter than the Gyr--averaged timescales discussed in this study (Sec.~\ref{discussion}). 

Throughout this paper the term `disk' is used to define this region (with a typical\footnote{The physical scales quoted here in kpc are only given as examples for typical M$_\star$ star--forming galaxies, and will scale as a function of the actual mass of a given dark matter halo. For a dependence of r$_{\rm stars}$ on r$_{\rm vir}$ see, e.g., \citet{somerville18}.} radius r$_{\rm stars}\,<$\,10\,kpc). Note that the term `disk' should not be taken literally: for example, low mass galaxies may not form well--defined disks, and many massive disk galaxies will transition to elliptical galaxies through mergers over time. We thus consider the `disk' nomenclature to define the main stellar components of galaxies, which, for main sequence star--forming galaxies at high redshift, can be considered disk--like in many cases \citep[e.g.,][]{foerster-schreiber09, wuyts11, salmi12, law12}. 

This nominal `disk' region is surrounded by a reservoir of atomic gas (\ion{H}{1}) with radii r$_{\rm HI}<$50\,kpc (Sec.~\ref{HI}), as demonstrated by observations in the local universe \citep[e.g.,][]{walter08, leroy09}, high redshift observations \citep{krogager17, neeleman17, neeleman19} as well as simulations \citep[e.g.,][]{bird14, rahmati14}. Outside the atomic gas region is the circumgalactic medium (CGM), defined to be located within the virial radius ($r_{\rm vir}\!\sim$\,50--300\,kpc), meaning gravitationally bound to the dark matter halo, and decoupled from the expansion of the Universe \cite[e.g., ][]{tumlinson17}. The CGM consists of predominantly ionized gas at a range of temperatures (T$\sim$10$^4$\,--10$^6$\,K). The timescale to accrete material from the cool $T\sim10^4$\,K CGM is comparable to the dynamical time ($\sim\,10^8$\,yr), orders of magnitudes shorter than the cooling time of the hot $T\sim10^6$\,K CGM  ($\gg\,1$\,Gyr, Sec.~\ref{accretion}).  The medium outside this gravitationally collapsed/bound structure (i.e., beyond $r_{\rm vir}$) is referred to as the intergalactic medium (IGM). 

The above defined regions are not static, and gas can be exchanged between these regions. The most important gas flows are also included in the schematic shown in Fig.~\ref{fig:schematic}, i.e. outflows as well as gas accretion. As detailed below (Sec.~\ref{rates}),  the accretion process can be described as: (i) the net infall of ionized material from the CGM and/or IGM onto the extended \ion{H}{1} reservoir, and (ii) the net inflow of \ion{H}{1} from the \ion{H}{1} reservoir (within r$_{\rm HI}$), with the subsequent conversion to H$_2$, onto the central region of the galaxy (within r$_{\rm stars}$). We also note that our schematic does not include the accretion of mass through galaxy mergers. Their contribution to the mass build--up in galaxies is significantly smaller than that from accretion \citep[e.g., ][]{vandevoort11}.

We emphasize that the demarcation of IGM versus CGM versus `disk' is not a simple geometric one, with material necessarily transitioning from one region to the other over  time. For instance, the \ion{H}{1} and warm/hot halo gas may mix substantially through streams, Galactic fountains and outflows, as well as filaments. Likewise, many galaxies reside in groups or clusters, where the dark matter halos may overlap, and defining whether gas is in the IGM vs.\ CGM may be ambiguous. However, for the purpose of the analysis presented in this paper, where we focus on the evolution of the baryonic components of the `disk' structure, the proposed simple schematic in Fig.~\ref{fig:schematic} should suffice as a representative guide.

\begin{figure*}
\centering 
\includegraphics[angle=0,width=1.6\columnwidth]{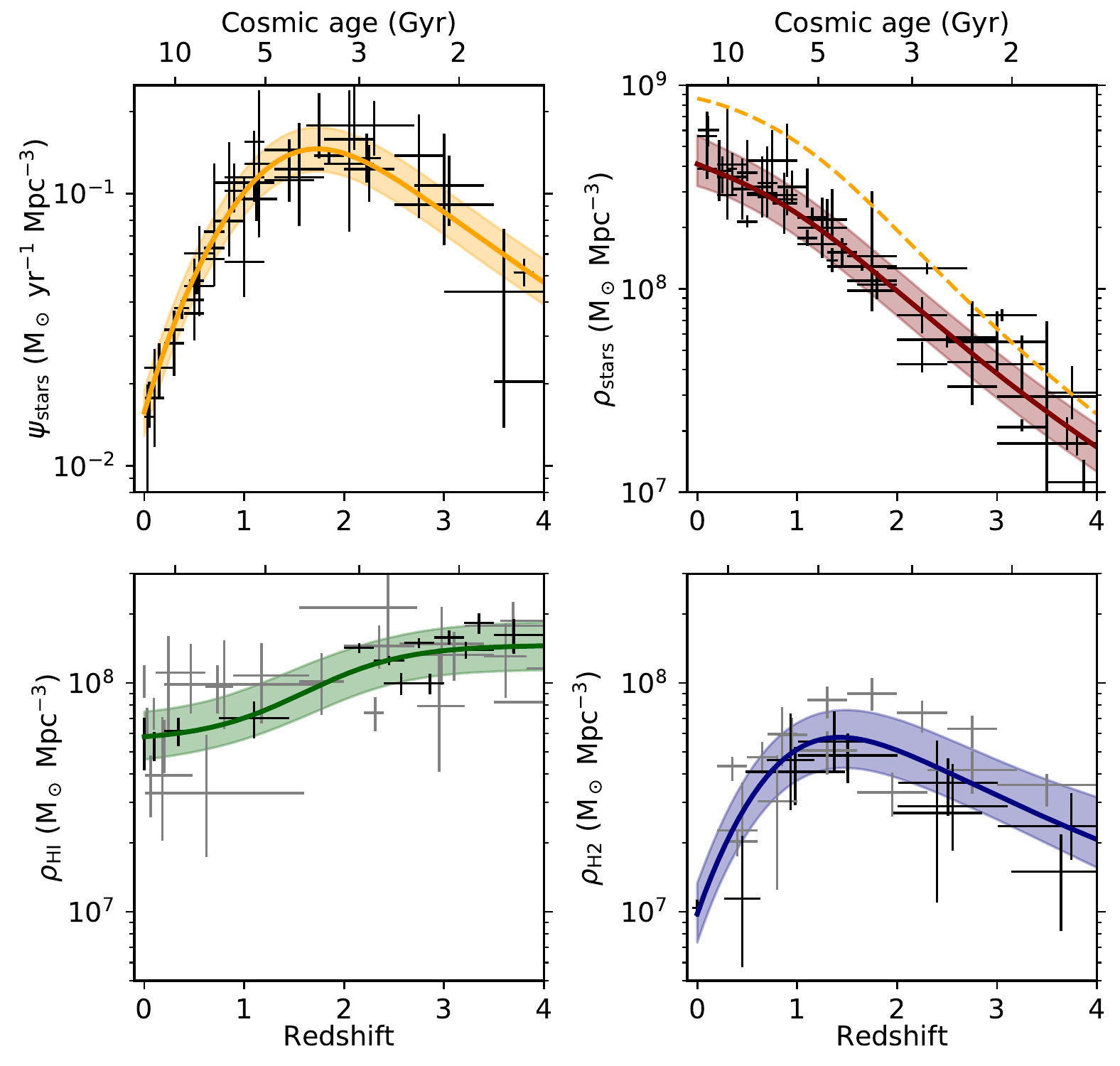}
\caption{Redshift evolution of different baryonic components in galaxies compiled from the literature. The measurements of the cosmic star formation rate density {\em (top left)} and stellar mass density {\em (top right)} are from the compilation in \citet{madau14} (their tables 1 and 2). The solid line is the best-fit functional form to the data (section \ref{sec:fitfun}) with the parameters given in Table \ref{tab:fitfun}, and the shaded region marks the 1$\sigma$ region (16$^{\rm th}$ to 84$^{\rm th}$ percentile) from a Monte Carlo Markov Chain analysis (see Sec.~\ref{sec:fitfun} for details). The orange dashed line in the cosmic stellar mass density panel is the integration of the best fit function form to the star formation rate density. The discrepancy between this curve and the measurements is described by the return fraction \citep[see text and][]{madau14}. Observational constraints on \rhohi\ {\em (bottom left)} are from a compilation given in \citet{neeleman16} updated with some recent constraints at low redshift (Sec.~\ref{HI}). Grey points indicate measurements at $<\,$6$\sigma$ and black points are measurements at $>\,$6$\sigma$ (see Appendix~B). Constraints on $\rho_{\rm H2}$ {\em (bottom right)} are from ASPECS \citep[][]{decarli19,decarli20} and other CO surveys (black points; see Appendix~B). Grey points indicate measurements obtained through dust continuum observations  \citep[Appendix~B; including the ASPECS measurements by][]{magnelli20}.
\label{fig:fitfun}}
\end{figure*}

\section{Mass components}
\label{baryons}

To put the different baryonic mass components in galaxies in context, we here compile current literature estimates of their `cosmic mean density' as a function of redshift. The total number of baryons is conserved over time, and therefore, by definition, the density of baryons does not change with time when considering co--moving volumes\footnote{Strictly speaking, the baryon density decreases  with time due to fusion, as some of the mass is converted to energy. E.g. in the case of the fusion of two hydrogen atoms to form Helium, 0.7\% of the mass is lost to radiation. During a complete CNO cycle, approximately the same amount of energy is being released. As only a small fraction of all baryons, those within the centers of stars, take part in the fusion process, we ignore this mass--loss here.}. 

\subsection{The $z\,\sim\,0$ census}
\label{redshift0}

For the low redshift Universe ($z\,\lesssim\,0.3$), an almost complete census of the baryons is available \citep[e.g.,][]{shull12, tumlinson17,nicastro18}. The latest studies place the large majority ($\sim$\,82\%) of the cosmic baryons in the IGM \citep[e.g.,][]{shull12}. These baryons are highly ionized (temperatures between $10^5$ and $10^7$\,K), and detected via \ion{O}{6} and \ion{O}{7} absorption features, and as the Ly--$\alpha$ forest. The distribution is thought to be highly filamentary, with the majority of the IGM residing in the `cosmic web'. Recent work on fast radio bursts has shown promise to detect this hard--to--trace component \citep[e.g.,][]{mcquinn14, shull18, macquart20}.

The remaining 18\% of the baryons at $z\,\approx\,0$ then belong to the `collapsed phase' \citep[][see also their Figure~10]{shull12}, gravitationally bound to galaxies, groups, and clusters that we will discuss in the following. The hot intercluster medium (ICM; $\ge 10^7$ K), seen in X--rays, comprises 4\% of the cosmic total\footnote{The ICM is not labeled in the schematic shown in Fig.~\ref{fig:schematic} as it only applies for cluster environments}. The stars in all types and masses of galaxies comprise 7\% of the total baryon density. The cold gas (\ion{H}{1} and H$_2$) comprises a little more than one percent at $z\,=\,0$, $\sim$85\% of which is in \ion{H}{1}. The CGM (also called `hot halos'), comprises about 5\% of the cosmic total, although again, the exact demarcation of the CGM remains somewhat subjective \citep[see also][and Sec.~\ref{schematic}]{shull12, tumlinson17}. 

There are other mass components in galaxies, but they only marginally contribute to the total mass budget, as briefly summarized in the following. As their combined contribution is of the order a few percent, we do not consider them further in our analysis.

\noindent {\em Warm ionized medium:} The warm ionized medium (WIM) is visible in H$\alpha$ and X--rays, and makes up less than 1\% of the total baryon mass in galaxy disks \citep[e.g.][]{anderson10, putman12, werk14}.

\noindent {\em Black holes:} The majority of galaxies are thought to host central supermassive black holes (SMBH). Various studies put this ratio at $\sim\,$0.1\% of the total stellar mass in galaxies \citep{kormendy13}. The remnants of massive stars are by definition included in the stellar Initial Mass Function (IMF) determinations \citep[e.g., IMF review by][]{bastian10}. 
Some black holes may be ejected entirely from galaxies via interactions with other black holes, but this net mass effect is minor \citep[e.g.][]{loeb07}.

\noindent {\em Dust:} Although dust plays a central role in the formation of stars, dust only makes up about 1\% of the total ISM mass \citep[e.g.,][]{sandstrom13}. The cosmic evolution of the dust content in the universe has recently been discussed in \cite{driver18} and \cite{magnelli20}.

\subsection{Redshift evolution}
\label{baryons_galaxies}

In the following we discuss the key baryonic mass components in galaxies, and their evolution with cosmic time.

\subsubsection{Star formation and Stars}
\label{stars}

The evolution of the cosmic star formation rate density ($\psi_{\rm stars}$; Fig.~\ref{fig:fitfun}, top left) has been constrained through various multi--wavelength studies of large samples of individual galaxies over the last decades \cite[as summarized in the review by][]{madau14}. Early studies were based on rest--frame UV observations \citep[e.g.,][see above review for a complete list of references]{madau96, lilly96, cucciati12, bouwens12a, bouwens12b}, and are complemented through observations at longer wavelengths
\citep[e.g.,][]{magnelli11, magnelli13, gruppioni13, sobral13, bouwens16, bouwens20, novak17, dudzevivciute20, khusanova20}. These estimates indicate that the peak of cosmic star formation occurred at $z\,\sim\,2$, with a subsequent decline by a factor of $\sim$\,8 to the present day. Integrating $\psi_{\rm stars}$ gives the stellar mass formed at a given cosmic time, and this integral is shown as a dashed orange line in Fig.~\ref{fig:fitfun} (top right). 

This integral can be compared to the independently measured stellar mass density $\rho_{\rm stars}$ (shown as a red line in Fig.~\ref{fig:fitfun}, top right). This stellar mass density has been determined by numerous studies \citep[e.g.,][]{perezgonzalez08, marchesini09, caputi11, ilbert13, muzzin13}, as compiled and homogenized in the review by \citet{madau14}. Both the stellar mass and the star formation rates depend on the choice of the IMF, and SED fitting method \citep[e.g.,][]{bastian10, kennicutt12, leja20}\footnote{\citet{madau14} assume a Salpeter IMF and a lower fixed threshold in luminosity of  0.03\,$L_\star$}.

This temporal integral of the star--formation rate density lies above the measured stellar mass density $\rho_{\rm stars}$ by a factor $1.4\,\pm\,0.1$. This is due to the fact that not all stellar mass that is formed will stay locked in stars; some fraction will be returned to the ISM, CGM, or IGM (depending on the mass of the galaxy). The cosmic--averaged star formation rate density $\psi_{\rm stars}(z)$ is thus the first time derivative of $\rho_\star(z)$, modulo the return fraction\footnote{The return fraction is $R$\,=\,0.27 for a Salpeter IMF and $R$\,=\,0.41 for a Chabrier IMF that is more weighted towards massive stars \citep{madau14}.} of stars $R$ to the interstellar medium through stellar winds and/or supernova explosions \citep[e.g.,][]{madau14}, i.e.,

\begin{equation*}\label{eq_rhosfr}
\dot \rho_{\rm stars}(z) = (1-R) \, \psi_{\rm stars}(z). 
\end{equation*}

The fact that the integral of $\psi_{\rm stars}$, after accounting for the return fraction, is in reasonable agreement with $\rho_{\rm stars}$ is remarkable, as highlighted in \citet{madau14}, if one considers the number of assumptions that go into each measurement\footnote{But see \citet{hopkins18} who argues that this overall agreement does not necessarily imply that the IMF has to be universal.}. These mass estimates do not include stars that are found outside galaxy disks, e.g. in stellar streams around galaxies, and the intracluster environment. This stellar mass component, however, only constitutes a small fraction of the stellar mass present in the galaxy disks \citep[e.g.,][]{behroozi13}, and we therefore do not consider this component further. For completeness it should be noted that some of the stellar mass growth can occur through mergers of galaxies, but this gain (`ex-situ') through merging of the existing stellar masses is small compared to the actual star formation process (`in--situ'), at least for galaxies around $L^\star$ \citep[e.g.][]{behroozi19}.

\begin{figure}
\centering 
\includegraphics[width=\columnwidth]{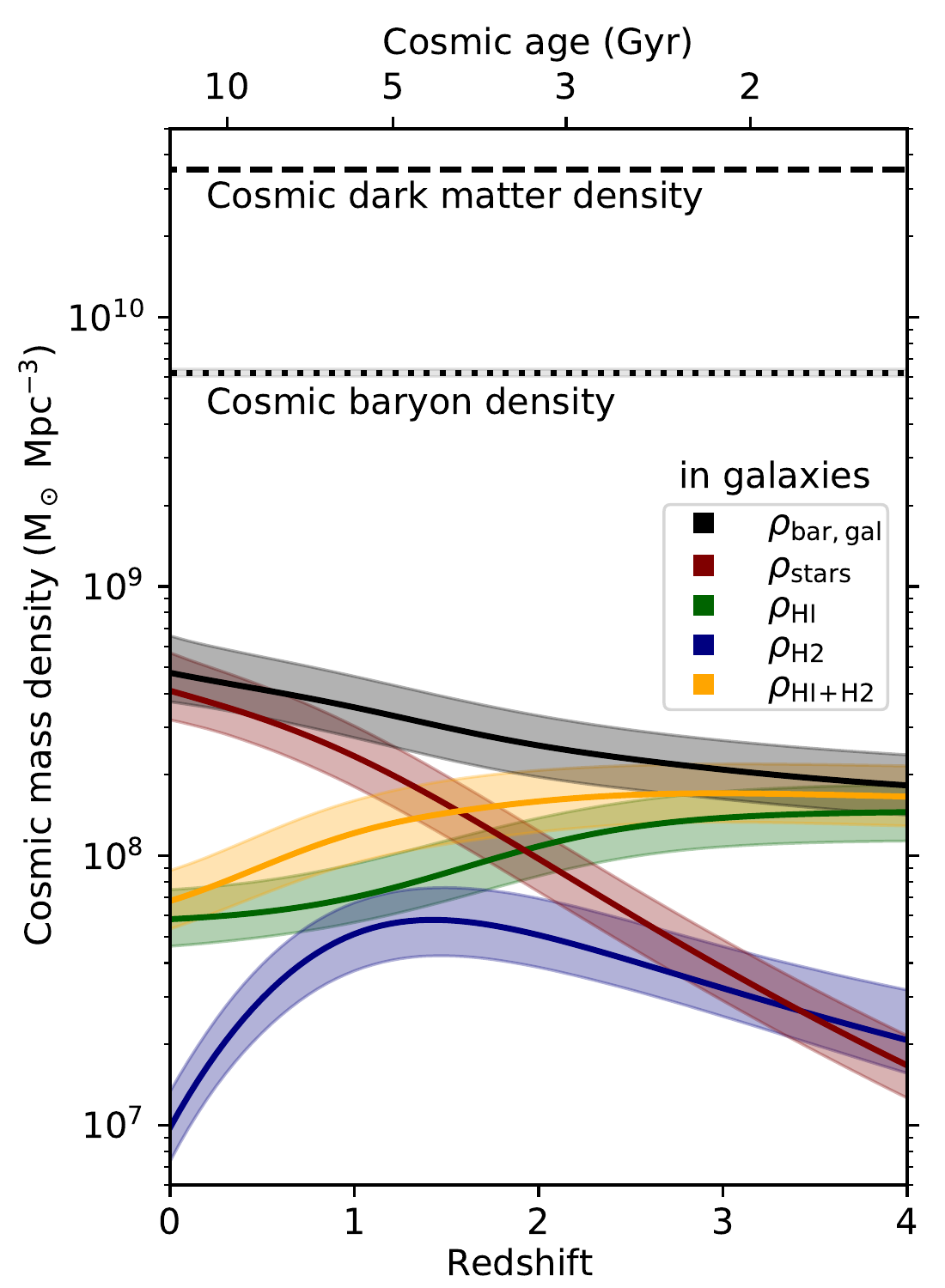}
\caption{Census of baryons inside and outside galaxies using the fitting functions shown in Fig.~\ref{fig:fitfun} and presented in Sec.~\ref{sec:fitfun}. Colors are as in Fig.~\ref{fig:fitfun}. The orange line shows the sum of the \HI\ and H$_2$ components, whereas the black line shows the sum of all of the baryons (stars, \HI\ and H$_2$) associated with galaxies. The dotted line is the total cosmic baryon content and the dashed line is the total dark matter content for the given $\Lambda$-CDM Universe. The same curves are plotted as a function of (linear) time in Fig.~\ref{fig:future}.
\label{fig:summary}}
\end{figure}

\subsubsection{Atomic gas}
\label{HI}

The evolution of the cosmic density of atomic gas associated with galaxies (\rhohi$(z)$; Fig.~\ref{fig:fitfun}, bottom left) has been constrained with 
several different approaches depending on redshift range. At $z\,\approx\,0$, large surveys aimed at measuring the \HI~21\,cm emission from local galaxies can constrain the \HI\ mass function \citep[e.g.,][]{zwaan05, braun12, jones18} whose integral provides an estimate of \rhohi. At higher redshifts ($0.3\,\lesssim\, z\,\lesssim\,1$), where the \HI~21\,cm emission becomes increasingly faint and therefore single sources are below the detection threshold of current radio--wavelength facilities, stacking of \HI~21\,cm emission from a large sample of galaxies provides an alternative approach to measure \rhohi$(z)$ \citep[e.g.,][]{lah07, delhaize13, rhee13, kanekar16, bera19}. In addition, the cross--correlation between 21\,cm intensity maps and the large scale structure (so--called 21\,cm intensity mapping) provides an independent measurement of \rhohi\ at these redshifts \citep[e.g.,][]{masui13, switzer13}. 

At $z\,\gtrsim\,1.6$, \HI\ can be observed using ground--based optical telescopes through its Ly$\alpha$ transition. Quasar absorption spectroscopy of the strongest Ly$\alpha$ absorbers, the so--called damped Ly$\alpha$ systems \citep[DLAs;][]{wolfe05} has yielded estimates of \rhohi\ up to $z\,\sim\,5$ \citep[e.g.,][]{crighton15}. The \rhohi\ estimate obtained from DLA surveys is simply the total \ion{H}{1} column density detected in DLAs divided by the path length of the survey. Here the main uncertainties come from relatively poorly understood systematics between varying methods of measuring DLAs and a potential bias against dusty, high \HI\ column density systems \citep{ellison01, jorgenson06, krogager19}. Most numerical simulations predict DLAs to probe gas near galaxies \citep{rahmati14}, which is supported by observations of the cross-correlation function between DLAs and the Ly$\alpha$ forest \citep{perez-rafols18}.

These measurements do not include any contributions from systems below the DLA column density threshold, because these systems contain less than 20\% of the total cosmic atomic gas density \citep{peroux03, omeara07, noterdaeme12, berg19}, and their connection with galaxies is less certain. However, we do account for the contribution of helium, which corresponds to a correction factor of $\mu = 1.3$.

The emerging picture is that the cosmic density of neutral atomic gas remains approximately constant with redshift, with a decline by a factor of $\sim$\,2 from $z\,\sim\,3$ to $z\,=\,0$. We remind the reader that the \HI\ is coming from a more extended reservoir compared to the stellar mass and star formation measurements of galaxies (see discussion in Sec.~\ref{schematic}).

\begin{figure*}
\centering 
\includegraphics[width=2.15\columnwidth]{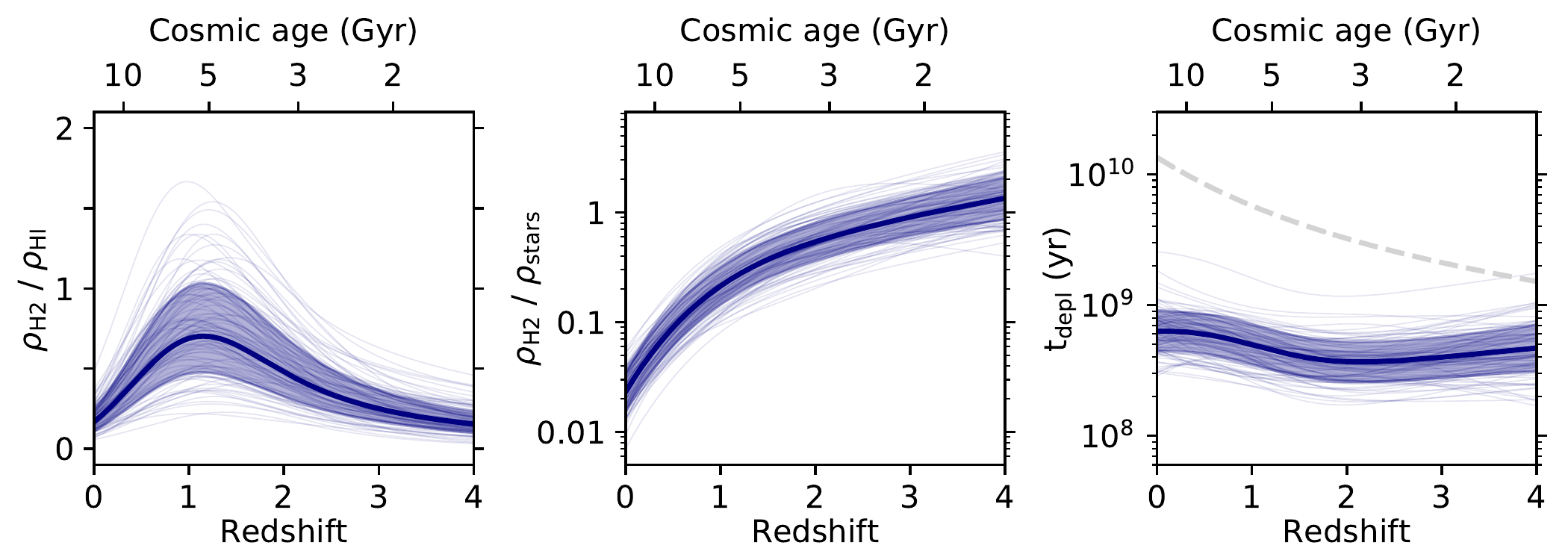}\\
\caption{{\em Left:} The ratio of cosmic molecular--to--atomic gas density as a function of redshift. The ratio peaks at $z\,\sim\,1.5$, close to the peak of the star formation rate density. {\em Middle:} The ratio of the molecular gas--to--stellar mass density as a function of redshift. {\em Right:} Cosmic gas depletion timescale, defined as the density in molecular gas divided by the cosmic star formation rate density. The grey dashed curve is the Hubble time vs.\ redshift. In all panels the thick solid line is derived from the functional form to the data (section \ref{sec:fitfun}) with the parameters given in Table \ref{tab:fitfun}. The shaded region marks the 1$\sigma$ region (16$^{\rm th}$ to 84$^{\rm th}$ percentile) of all the curves from a Monte Carlo Markov Chain analysis. Thin lines show several random realizations of this analysis.
\label{H2division}}
\end{figure*}

\subsubsection{Molecular gas}
\label{H2}

A number of approaches have been followed in the past to constrain the evolution of the cosmic molecular gas density ($\rho_{\rm H2}$; Fig.~\ref{fig:fitfun}, bottom right). Here we focus on methods that are not merely based on stellar mass and star--formation rate determinations with subsequent application of scaling relations. In particular, we here include the recent results from ASPECS, that perform deep frequency scans to detect redshifted CO lines without any pre--selection. This  approach has been successfully applied in a number of studies \citep{walter14, walter16, decarli14, decarli16, decarli19, decarli20, riechers19, pavesi18, klitsch19, lenkic20}. Molecular gas constraints derived from  dust emission (frequently using scaling relations based on stellar mass or star formation rates) and other approaches show a consistent evolution \citep[e.g.,][]{berta13, scoville17, driver18, liu20, magnelli20,dudzevivciute20}.

In order to convert CO to H$_2$ measurements, the detected CO emission has to be corrected for excitation and a CO--to--H$_2$ conversion factor has to be applied (that also accounts for helium). The  CO--to--H$_2$ conversion factor is the main systematic uncertainty in the analysis. For the ASPECS measurement, the majority of the molecular gas mass density comes from individually detected galaxies \citep{decarli19}. Their metallicities (consistent with solar) and stellar masses (M$_{\rm stars} \gtrsim 10^{10}$\,M$_\odot$) justify the choice of a Galactic conversion factor to determine the molecular gas mass \citep{boogaard19}. The uncertainties in molecular gas excitation, as derived for the ASPECS galaxies in \citet{boogaard20}, have been anchored based on CO(1--0) observations out to $z\,\sim\,3$ \citep[VLASPECS,][]{riechers20}, and were folded into the ASPECS measurements \citep{decarli20}. Converting dust measurements to molecular gas masses requires the choice of a dust temperature, emissivity, and a dust--to--gas ratio (see above references).

Stacking and intensity mapping techniques \citep{inami20, uzgil20} indicate that the majority of all CO emission in the UDF is captured by the current observations, i.e. the faint--end slope of the CO luminosity functions is such that extrapolating to lower masses would not significantly (less than 50\%) increase the total emission \citep[see also][]{decarli20}. These high--redshift measurements are anchored at $z\,=\,0$ through detailed studies of the molecular gas content in the local universe \citep{keres03, boselli14, saintonge17, fletscher20}.

The emerging picture based on the above--mentioned molecular gas and dust studies is that the cosmic density of molecular gas decreased by a factor of $6^{+3}_{-2}$ from the peak of cosmic star formation ($z\,\sim\,2$) to today \cite[see also recent reviews by][]{tacconi20,peroux20,hodge20}. There is evidence that the molecular gas density increased from $z\,\sim\,6$ to $z\,\sim\,2$ \citep{riechers19, decarli19, decarli20}, but the associated uncertainties are significant for $z\,>\,3$.

\begin{deluxetable*}{lccccc}
\tablecaption{Fitting functions to the observed cosmic density measurements shown in Fig.~\ref{fig:fitfun}
\label{tab:fitfun}}
\tablehead{
\colhead{} & 
\colhead{Fitting function} &
\colhead{$A$} &
\colhead{$B$} &
\colhead{$C$} &
\colhead{$D$}} 
\startdata
$\rho_{\rm H_2}$($z$)[$M_\odot$ Mpc$^{-3}$] & Equation~1 & $(1.00 \pm 0.14) \times 10^7$ & $3.0 \pm 0.6$ & $2.3 \pm 0.3$ & $5.1 \pm 0.5$\\
$\rho_{\rm stars}$($z$)[$M_\odot$ Mpc$^{-3}$] & Equation~1 & $(1.3^{+1.0}_{-0.6}) \times 10^{10}$ & $ -4.1 \pm 0.4$ & $2.5 \pm 0.4$ & $-3.8 \pm 0.3$ \\
$\psi_{\rm stars}(z)$[$M_\odot$ yr$^{-1}$ Mpc$^{-3}$] & Equation~1 & $0.0158 \pm 0.0010$ & $ 2.88 \pm 0.16$ & $2.75 \pm 0.11$ & $5.88 \pm 0.15$\\
\rhohi($z$)[$M_\odot$ Mpc$^{-3}$] &  Equation~2 & $(4.5 \pm 0.5) \times 10^7$ & $2.8 \pm 0.4$ & $(1.01 \pm 0.07) \times 10^8$ & ---\\
\enddata
\end{deluxetable*}

\subsection{Fitting functions}
\label{sec:fitfun}

In order to capture the global trends in the cosmic density measurements discussed in the previous paragraphs, we have fitted the observational data with functional forms (data given in Appendix B). In particular, for $\rho_{\rm H2}$, $\rho_{\rm stars}$, and $\psi_{\rm stars}$, we adopt a smooth double power law, similar to that defined in \citet{madau14}:

\begin{equation}
\rho_x(z) = \frac{A (1+z)^B}{1 + [(1 + z)/C]^D}
\end{equation}

In order to capture the apparent flattening of the evolution of \rhohi\ at both low and high redshift, we adopted a hyperbolic tangent function \citep[as in][]{prochaska18}

 \begin{equation}
 \rho_{\rm H{\scriptscriptstyle I}}(z) = A \tanh(1 + z - B) + C
 \end{equation}
 
These functional forms are not physically motivated and are simply meant to capture the general trends of the data points. To estimate the best fit parameters and associated uncertainties, we fit the data using a Monte Carlo Markov Chain approach utilizing the \emph{emcee} package \citep{foremanmackey13}. For all cosmic densities, we marginalize over a nuisance parameter to account for intrinsic scatter within the data points not accounted for by the uncertainties of the individual points. To take into account systematic uncertainties within the varying data sets, we symmetrically (in log--scale) increase the formal uncertainties derived from the fitting procedure such that $>$68\% of all measurements are contained within the 1$\sigma$ boundaries (16$^{\rm th}$ to 84$^{\rm th}$ percentile). The best fits are shown as solid lines in Figs.~\ref{fig:fitfun} and~\ref{fig:summary}, whereas the 1$\sigma$ boundaries of the fitting functions are shown as colored regions. The fitting parameters are summarized in Table~\ref{tab:fitfun}. We note that a fit to $\rho_{\rm H2}$ based on just the ASPECS data gives almost identical parameters as those shown in Table~\ref{tab:fitfun}.

\subsection{Cosmic Averages}
\label{averages}

In the analysis that follows, we will consider the above volume--averaged measurements (Sec.~\ref{baryons_galaxies}) to derive volume--averaged properties (such as depletion times, gas accretion rates). The fundamental assumption is that, statistically speaking, the galaxies are similar to the picture discussed in Sec.~\ref{schematic} and Fig.~\ref{fig:schematic}. One can express the quantities discussed here as a function of the well--characterized stellar mass function (SMF) $\Phi_\star(z,M)$ \citep[e.g.][]{davidzon17}. Then the cosmic stellar mass density can be written as:

\begin{equation*}
\label{eq_rhostar}
\rho_\star(z)= \int\Phi_\star(z,M_\star) dM_\star,
\end{equation*}
where $M_\star$ is the stellar mass.
The gas (H$_2$ or \ion{H}{1}) density can then be expressed as:

\begin{equation*}\label{eq_rhoh2}
\rho_{\rm gas}(z)=\int\Phi_\star(z,M_\star)\times f_{\rm gas}(z,M_\star) \, dM_\star,
\end{equation*}
where $f_{\rm gas}$ is the gas--to--stellar mass fraction ($f_{\rm H2}$ or $f_{\rm HI}$).

By definition, these functions are volume averages that marginalize over dependencies of baryonic components on other parameters (such as, e.g., environment, metallicity, feedback processes).

\section{Discussion}
\label{discussion}
We now discuss the density evolution of the various mass components in the Universe and implications for gas accretion rates. As stressed before, our measurements are volume-- and time--averaged. The timescales of the individual mass conversion processes ($\lesssim$0.1\,Gyr, Sec.~\ref{schematic}) are smaller than the cosmic timescales over which we are averaging ($\Delta z\,=\,1$ corresponds to a time period of $\sim$\,0.6\,Gyr at $z\,=\,3.5$, $\sim$\,2.5\,Gyr at $z\,=\,1.5$, and $\sim$\,5.5\,Gyr at $z\,=\,0.5$). Therefore, our conclusions will not be applicable to all individual galaxies.

\subsection{The evolution of the cosmic baryon density}

Fig.~\ref{fig:summary} summarizes the evolution of the baryon content in stars, \ion{H}{1}, and H$_2$ associated with galaxies together with the cosmic dark matter and the total baryon density. As discussed in Sec.~\ref{baryons}, the large discrepancy between the total baryon density (dotted curve in Fig.~\ref{fig:summary}) and the baryon density inside galaxies $\rho_{\rm bar, gal}$ (black curve in Fig.~\ref{fig:summary}) indicates that most baryons are not inside galaxies, but are in the predominantly ionized IGM (and CGM). The stellar mass density is increasing continuously with time, and surpasses that of the total gas density (\ion{H}{1} and H$_2$) at redshift $z\,\sim\,1.5$.

In Fig.~\ref{H2division} (left) we plot the ratio of molecular to atomic gas density as a function of redshift. This ratio peaks at $z\,\sim\,1.5$, close to the peak of the star formation rate density. Fig.~\ref{H2division} (middle) shows the ratio of molecular gas--to--stellar mass as a function of redshift. At redshifts $z\,\lesssim\,2$ the stellar mass density starts to dominate over the molecular gas density.

The last panel in Fig.~\ref{H2division} (right) shows the molecular gas depletion time, i.e., how long will it take to deplete the molecular gas reservoir at the current rate of star formation. The depletion time ($\rho_{\rm H2}$/$\psi_{\rm stars}$) is approximately constant above redshifts $z\,\gtrsim\,2$, and then increases slightly from $\tau_{\rm depl}\,\sim\,(4 \pm 2)\,\times\,10^{8}$\,yr at $z\,\sim\,2$ to $\tau_{\rm depl}\,=\,(7 \pm 3)\,\times\,10^{8}$\,yr at $z\,=\,0$, and is shorter than the Hubble time at all redshifts. This immediately implies that the molecular gas reservoir needs to be continuously replenished (i.e., through accretion). Both the ratio of molecular gas--to--stellar mass and the depletion times for the molecular gas phase are similar to what is found in scaling--relation studies of individual galaxies \citep[e.g.,][]{daddi10, genzel10, bothwell13, tacconi18, aravena19, aravena20}.

\subsection{The need for accretion}
\label{sec:accretion}

The need for gas accretion onto galaxies from the cosmic web to sustain the observed star formation activity has been noted numerous times before \citep[e.g.,][]{bouche10,bauermeister10,dave12,lilly13,conselice13,bethermin13,behroozi13,tacconi13,peng14, rathaus16, scoville17, tacconi18}. Prior to the availability of direct measurements of the H$_2$ density it was occasionally argued that, given the approximate constancy of the \ion{H}{1} density through cosmic time, the net gas accretion rate density needed to be approximately equal to the star formation rate density. Now that the molecular density is directly observed, this topic can be revisited \citep[see also the recent reviews by][]{peroux20,tacconi20,hodge20}. 

\begin{figure}
\centering 
\includegraphics[width=0.5\textwidth]{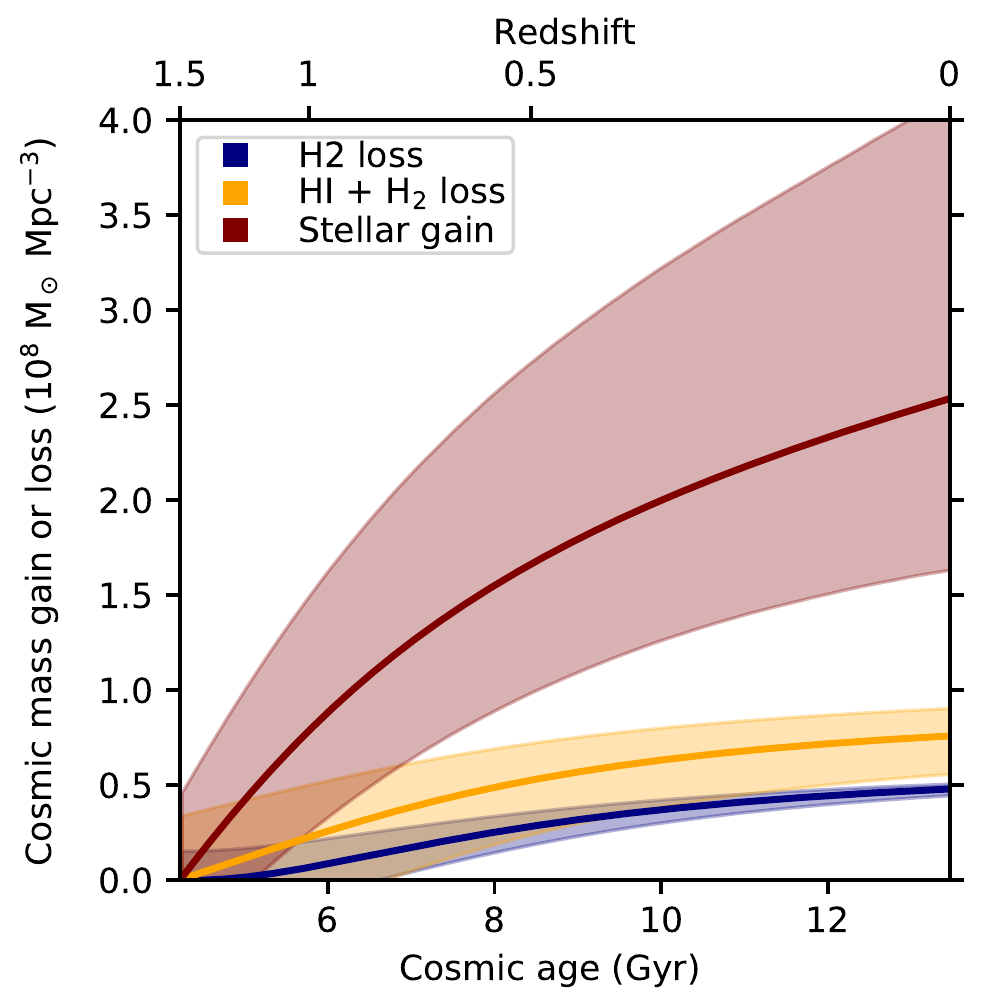}
\caption{Cumulative {\em gain} of the stellar mass density (red line) compared to the cumulative {\em loss} of the gas mass density (H$_2$: blue line, total gas: orange line), starting at a redshift of $z\,=\,1.5$ ($T_{\rm Univ}\!\sim4$\,Gyr), i.e., approximately the peak of the molecular gas density. The lower (upper) abscissa shows cosmic age (redshift). Even assuming  that 100\% of the gas will end up in stars, the gas observations cannot account for the observed stellar mass build--up. The remaining mass to build up the stellar mass must be accreted onto the galaxy.
\label{massloss}}
\end{figure}

We first ask how much stellar mass could in principle be formed by looking at the decrease in the molecular gas density since the peak of the cosmic molecular gas density at $z\,\sim\,1.5$. If we assume that the net loss in H$_2$ since that time is fully due to the formation of stellar mass, we can derive the maximum stellar mass growth due to this conversion. This is shown in Fig.~\ref{massloss} as the blue curve. For completeness, we also show the loss in \ion{H}{1} (orange line: sum of \ion{H}{1} and H$_2$ loss) that eventually may also end up as stellar mass over the same cosmic time via a transition through the molecular gas phase (Sec.~\ref{rates}). We compare this to the  total observed gain in stellar mass over the same cosmic time (shown as the red curve in Fig.~\ref{massloss}, based on the red curve show in Fig.~\ref{fig:fitfun}). Even assuming that all the molecular gas ends up in stars, the observed decline in \ion{H}{1} and H$_2$ is only able to account for $\lesssim\,$25\% of the total stellar mass formed during this time. Also note that the above stellar mass measurement ignores the return of stellar mass to the ISM, CGM, and IGM. If this additional stellar mass is accounted for, the observed \ion{H}{1} and H$_{2}$ can only account for $\lesssim\,$20\% of the total stellar mass formed. The difference in mass is thus the minimum amount of material that needs to be accreted by the galaxies from the IGM/CGM since the Universe was 4\,Gyr old.

\subsection{\ion{H}{2}  infall and \ion{H}{1}  inflow rates}
\label{rates}
Most of the stars are thought to form out of H$_2$ and not atomic hydrogen \citep[e.g.,][]{schruba11}, at least at the redshifts considered in this paper. However, the presence of \ion{H}{1} is a prerequisite to form H$_2$. In nearby galaxies it is found that \ion{H}{1}\ is significantly more extended than the stellar component, which also harbors most of the star formation and the H$_2$ \citep{walter08, leroy09}. At high redshift, the situation is likely very similar, as indicated by the fact that the impact parameter for the DLAs found in quasar spectra are $\le$\,50\,kpc (Sec.~\ref{HI}), whereas the stellar components are typically $\le$\,10\,kpc in size \citep[e.g.,][]{fujimoto17, elbaz18, jimenez-andrade19}. The fact that DLAs show little to no Lyman--Werner absorption from molecules also points towards the fact that the \ion{H}{1} is more extended than the H$_2$, i.e. that the DLAs contain negligible molecules \citep{noterdaeme08, jorgenson14, muzahid15}.

We here consider the accretion of material to the central star--forming `disk' as a two--step process. The first is the {\em net infall} of ionized gas (\ion{H}{2}) onto the extended \ion{H}{1} reservoir, $\widetilde{\psi}_{\rm HII\rightarrow HI}$, e.g. through cold--mode accretion (Sec.~\ref{theory}). In a second step the gas further cools and settles in the central region where it forms H$_2$, which we refer to as {\em net inflow}, $\widetilde{\psi}_{\rm HI\rightarrow H2}$. We stress that we can only consider {\em net} rates: it is also possible that H$_2$ (or \ion{H}{1}) is dissociated / photo--ionized to form \ion{H}{2} through feedback processes. Our data do not allow us to differentiate between inflows and outflows, and we here define the net flow rates in such a direction that they are likely positive, i.e. $\widetilde{\psi}_{\rm HII  \rightarrow HI}$\,$>\,0$, $\widetilde{\psi}_{\rm HI\rightarrow H2}$\,$>\,0$.  We note that strictly speaking we refer to net flow {\em rate densities} (averaged over cosmic volume) throughout this work. For simplicity, we however refer to these as {\em rates} throughout.

\begin{figure}
\centering 
\includegraphics[width=\columnwidth]{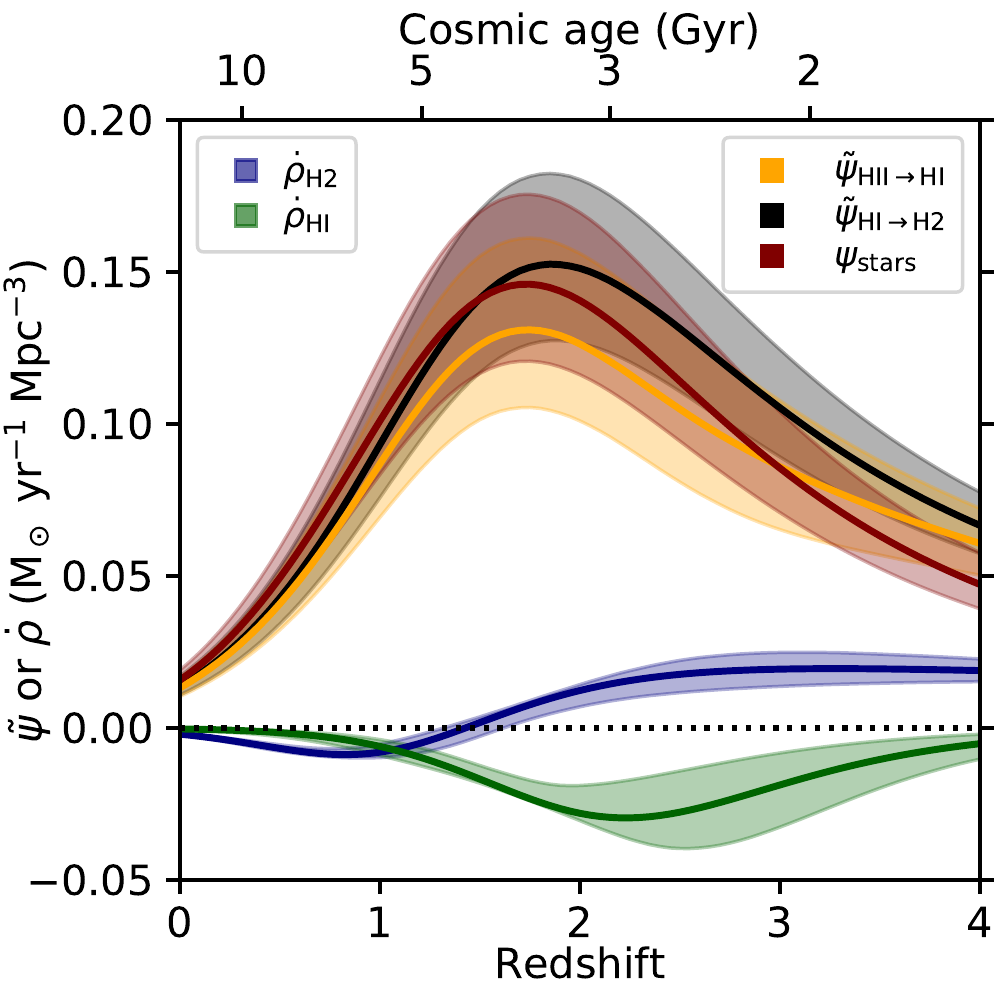}
\caption{The \ion{H}{2} net infall rate ($\widetilde{\psi}_{\rm HII\rightarrow HI}$, Eq.~\ref{eq_rhodotacc}, orange curve) and \ion{H}{1} inflow rate ($\widetilde{\psi}_{\rm HI \rightarrow H2}$, Eq.~\ref{eq_rhodotinf}, black curve) are plotted together with the cosmic star--formation rate density ($\psi_{\rm stars}$, red curve), assuming a mass loading factor of $\xi\,=\,0$. When including feedback / mass loading (i.e. $\xi\,>\,0$), the inflow and accretion rate would have to increase correspondingly, to account for the extra loss of gas.  We also show the time derivatives of the \HI\ and H$_2$ densities ($\dot \rho_{\rm HI}(z)$ and $\dot \rho_{\rm H2}(z)$), as derived from the temporal gradients of the measured density curves in Figure 2, as parameterized in equations 1 and 2. The curves of $\widetilde{\psi}_{\rm HI \rightarrow H2}$ and  $\widetilde{\psi}_{\rm HII\rightarrow HI}$ are a linear combination of the measured quantities: $\psi_{\rm stars}$, $\dot \rho_{\rm HI}(z)$, and $\dot \rho_{\rm H2}(z)$, as per equations 5 and 6. Below $z\,\approx\,1.5$ the inflow rate $\widetilde{\psi}_{\rm HI \rightarrow H2}$ drops below $\psi_{\rm stars}$, as the cosmic H$_2$ reservoir is used up to form stars (negative $\dot \rho_{\rm H2}(z)$). 
\label{fig:flows}}
\end{figure}

As detailed in Appendix~A, the rate at which the observed H$_2$ density $\rho_{\rm H2}$  is used up for star formation, lost due to feedback (both stellar or AGN) to the CGM, and is being replenished by \ion{H}{1} can be written as:

\begin{equation}\label{eq_rhodot_h2}
\dot \rho_{\rm H2}(z)=- \underbrace{\psi_{\rm stars}(z)}_{\substack{\text{star formation}\\\text{rate}}} - \underbrace{\xi\,\psi_{\rm stars}(z)}_{\substack{\text{H2 loss due}\\\text{to feedback}}}
+ \underbrace{\widetilde{\psi}_{\rm HI\rightarrow H2}(z)}_{\substack{\text{H2 gain from}\\\text{HI reservoir}}}
\end{equation}
where $\psi_{\rm stars}$ is the star formation rate density, and $\widetilde{\psi}_{\rm HI\rightarrow H2}$ is the net conversion rate of \ion{H}{1} to H$_2$; $\dot \rho_{\rm H2}(z)$ is the time derivative of $\rho_{\rm H2}(z)$. The (unknown) mass loading factor $\xi$ accounts for mass loss due to outflow driven by active star formation and AGN activity that is a function of the environment and mass distribution(s) within a galaxy. We here simplistically assume that this ouflow/mass loading is linearly correlated with the star formation rate density, $\psi_{\rm stars}$ \citep[e.g.,][]{spilker18, schroetter19} with a universal proportionality factor $\xi$.

The material that is required to replenish the \ion{H}{1} reservoir ($\dot \rho_{\rm HI}(z)$ being the time derivative of $\rho_{\rm HI}(z)$) can be expressed as:
\begin{equation}\label{eq_rhodothi}
\dot \rho_{\rm HI}(z)=-\underbrace{\widetilde{\psi}_{\rm HI \rightarrow H2}(z)}_{\text{loss to H2}} + \underbrace{\widetilde{\psi}_{\rm HII \rightarrow HI}(z)}_{\substack{\text{net HI gain from}\\\text{ HII reservoir}}}
\end{equation}
where $\psi_{\rm HII \rightarrow HI}$ is the net infall of gas from the ionized gas phase. As described in Appendix~A this expression for $\dot \rho_{\rm HI}(z)$ (unlike the one for $\dot \rho_{\rm H2}(z)$) does not include a mass loading term, as it is included in the net flow term $\widetilde{\psi}_{\rm HII \rightarrow HI}$.

We can solve equations ~\ref{eq_rhodot_h2} and~\ref{eq_rhodothi} for the net inflow rate $\widetilde{\psi}_{\rm HI \rightarrow H2}$ and the net infall rate $\widetilde{\psi}_{\rm HII\rightarrow HI}$ as a function of observables $\rho_{\rm HI}(z)$, $\rho_{\rm H2}(z)$ and $\psi_{\rm stars}$:

\begin{equation}\label{eq_rhodotinf}
\widetilde{\psi}_{\rm HI \rightarrow H2} = {\dot \rho_{\rm H2}(z)} + (1+\xi)\,\psi_{\rm stars}(z)
\end{equation}
and
\begin{equation}\label{eq_rhodotacc}
\widetilde{\psi}_{\rm HII\rightarrow HI} = 
{\dot \rho_{\rm HI}(z)} + 
{\dot \rho_{\rm H2}(z)} + (1+\xi)\,\psi_{\rm stars}(z).
\end{equation}

In Fig.~\ref{fig:flows} we plot these net flows rates (equations~\ref{eq_rhodotinf} and \ref{eq_rhodotacc}), along with the star formation rate density $\psi_{\rm stars}$. We also show the time derivatives of $\rho_{\rm HI}$ and $\rho_{\rm H2}$, derived as the proper time derivatives of the measured relations with redshift, as parameterized in Equations~1 and ~2. The differences between the net flow rates and the star formation rate density are due to the building up, or depletion, of gas in the neutral atomic and molecular phase, as dictated by the time derivative curves. 

At high redshift ($z\,>\,3$), both the net \ion{H}{2} infall rate ($\widetilde{\psi}_{\rm HII\rightarrow HI}$) and \ion{H}{1} inflow rate ($\widetilde{\psi}_{\rm HI \rightarrow H2}$) are larger than the star formation rate density, which is reflected in the build up of molecular gas over time, with the HI being a pass-through phase (close to zero derivative). At $z\,\gtrsim\,1.5$ the net {\em inflow} rate $\widetilde{\psi}_{\rm HI \rightarrow H2}$ is higher than $\psi_{\rm stars}$. At these redshifts, the H$_2$ cosmic density is still increasing with time. Therefore on top of the flow of H$_2$ into stars, additional accretion is needed to build up $\rho_{\rm H2}$, while HI is slowly being depleted. Conversely, at $z\,\lesssim\,1.5$ the net {\em inflow} rate $\widetilde{\psi}_{\rm HI \rightarrow H2}$ is lower than $\psi_{\rm stars}$. This is because the H$_2$ reservoir is decreasing with time in this redshift range, and therefore less H$_2$ needs to be replenished. 

\begin{figure*}
\centering 
\includegraphics[width=0.8\textwidth]{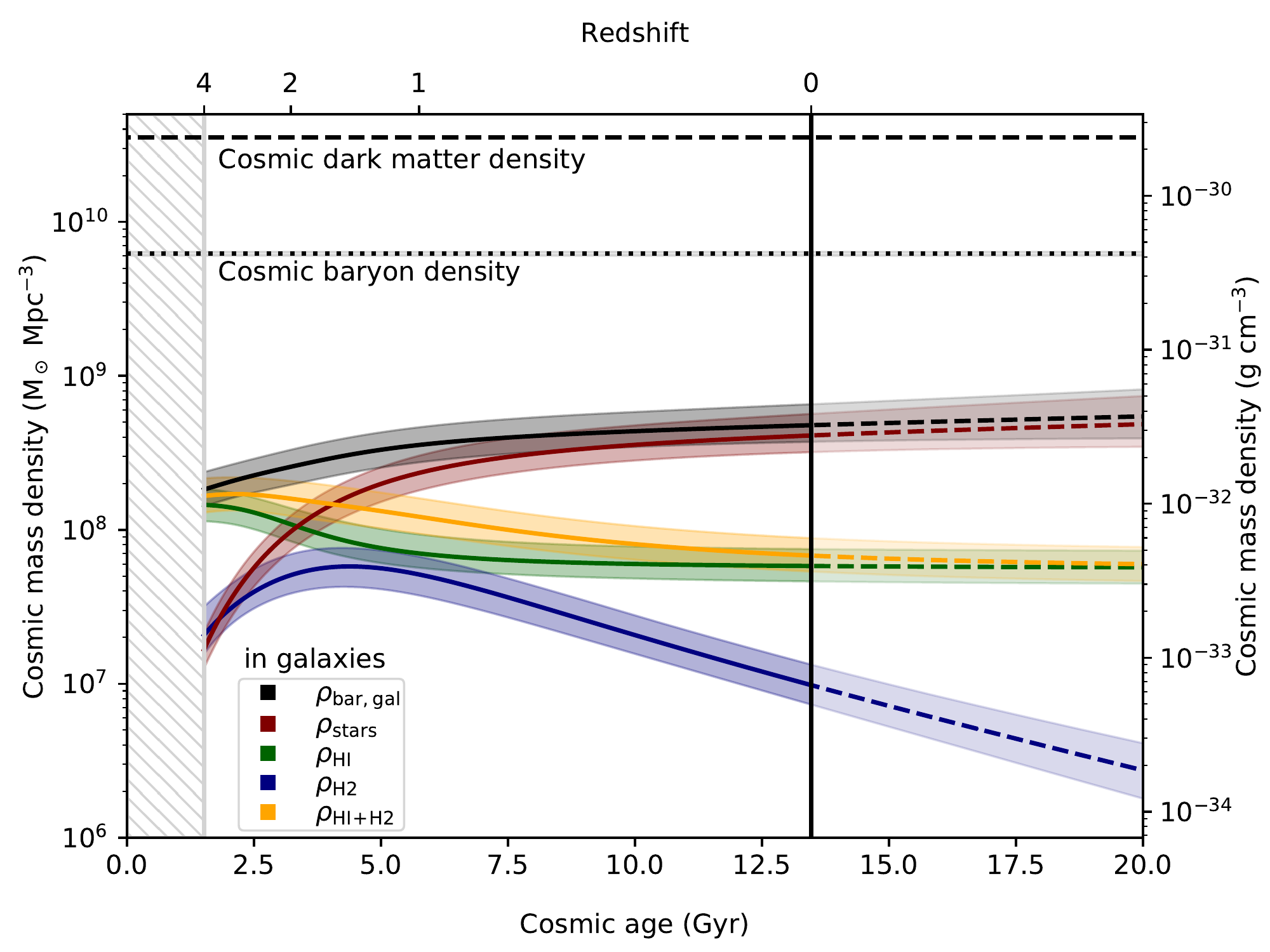}
\caption{We here plot the same information as in Fig.~\ref{fig:summary}, but with the following changes: (a) the lower abscissa shows cosmic time on a linear scale (redshift on the upper abscissa), (b) we extrapolate our fitting functions to the future (the present day is indicated by a vertical line, $z\,=\,0$), (c) we add units on the ordinate axis in g\,cm$^{-3}$. As in all other plots  we start plotting our functions at $z\,=\,4$. Under the assumption that our extrapolations are valid, the molecular gas density will decline by about a factor two over the next 5\,Gyr, the stellar mass will increase by approximately 10\%, and the inflow and accretion rates will decline correspondingly.}
\label{fig:future}
\end{figure*}

\subsection{The Cosmic Future}

Under the assumption of continuity, and that the physical process currently in play continue to dominate, we can use our empirical fitting functions (Sec.~\ref{sec:fitfun}) to forecast the evolution of the baryon content associated with galaxies over the next few Gyr. This is shown in Fig.~\ref{fig:future} where we plot the same information as in Fig.~\ref{fig:summary} but as a function of (linear) cosmic time. Assuming that our fits can be extrapolated to the future, the molecular mass density will decrease by about a factor of two over the next 5\,Gyr, the \ion{H}{1} mass density will remain approximately constant, and the stellar mass density will increase by about 10\%. The star--formation rate density will follow the decrease of H$_2$. Consequently, the total cold gas content in galaxies will be dominated by diffuse atomic gas even more than today. In this scenario, the ionized gas in the ICM/CGM will stay in this state and will not enter the main body of the galaxies. The inflow and infall rates (Eqs.~\ref{eq_rhodotinf} and \ref{eq_rhodotacc}) will decrease correspondingly.  Fig.~\ref{fig:future} shows that the Universe has entered `Cosmic Twilight', during which the star--formation activity in galaxies inexorably declines, as the gas inflow and accretion shuts down \citep[see also][]{salcido18}. Over this same time period, the majority of stars with masses greater than the Sun will have exceeded their main sequence lifetimes, leaving increasingly cooler, low mass stars to illuminate the Universe.

\subsection{Theory connection}
\label{theory}
Thus far, we have taken a strictly phenomenological approach to the trends observed in the data. We now discuss if cosmological simulations provide a sufficient amount of (dark and baryonic) matter to be accreted onto galaxy halos, to account for the observed net flows ($\widetilde{\psi}_{\rm HII \rightarrow HI}$ and $\widetilde{\psi}_{\rm HI \rightarrow H2}$). We also consider the potential role of preventive feedback mechanisms (such as virial shocks, AGN feedback, and cosmic expansion).

\subsubsection{Accretion onto dark matter halos}
\label{theory_accretion}

\begin{figure}
\centering 
\includegraphics[width=\columnwidth]{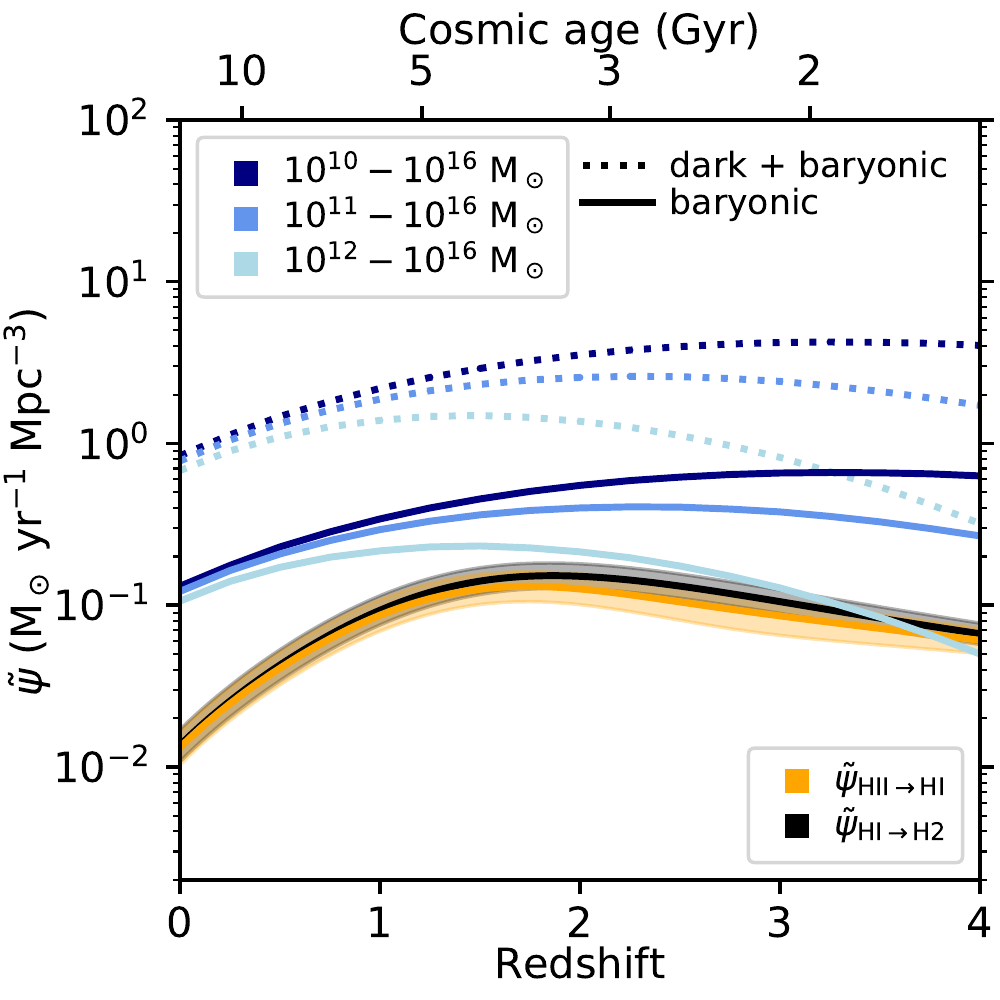}
\caption{Comparison of the observed net accretion rates (orange/black curves) and predictions from theory (blue curves). The observed net infall and net inflow rates onto the central disk are the same as in Fig.~\ref{fig:flows}, but are shown here on a logarithmic ordinate axis. The predictions from theory, based on the Bolshoi--Planck and MultiDark--Planck $\Lambda$CDM cosmological simulations, of the accretion rate of the total (dark and baryonic) matter onto the dark matter halo are shown as dashed blue curves for different virial mass ranges. The solid blue curves show the accretion rates for baryonic matter only (see discussion in Sec.~\ref{theory}). The predicted baryonic accretion rates  onto the galaxy halos are larger than the observationally required net infall rates  onto the central disk, indicating that most of the accreted baryons do not end up in the centers of galaxies.
\label{fig:theory}}
\end{figure}

We estimate the amount of baryonic matter that is accreted onto galaxy halos using the results from cosmological simulations. More specifically, we estimate the matter (dark and baryonic combined) accretion rate onto halos $\dot{M}_{\rm matter}(M_{\rm vir},z)$ as a function of halo virial mass and redshift using the fitting function presented in \citet[their equation 11 adopting the dynamically-averaged scenario]{RP2016}. The authors obtained this fitting function by measuring the growth of halos in the Bolshoi--Planck and MultiDark--Planck $\Lambda$CDM cosmological simulations \citep{Klypin2016}. The cosmic (dark + baryonic) matter accretion rate $\psi_{\rm matter}(z)$ is then obtained by taking the integral (over the virial masses considered) of the product between the matter accretion rate $\dot{M}_{\rm matter}(M_{\rm vir},z)$   and the number density of halos with that mass $\Phi_{\rm vir}(M_{\rm vir},z)$, such that

\begin{equation}
\psi_{\rm matter}(z) = \int \limits_{M_{\rm vir,min}}^{M_{\rm vir,max}} \hspace{-0.4cm} \dot{M}_{\rm matter}(M_{\rm vir},z)\times \Phi_{\rm vir}(M_{\rm vir},z) dM_{\rm vir}, 
\end{equation}
where the number density of halos as a function of virial mass and redshift is from equation 23 in \citet{RP2016}. These accretion curves are shown in Fig.~\ref{fig:theory} as dashed lines. The different lines show the accretion rates assuming different dark matter halo mass ranges, where the lowest mass considered here (M$_{\rm vir}\,=\,10^{10}$\,M$_\odot$) corresponds to the  mass resolution in the simulations considered (corresponding to a stellar mass of a few times $10^7\,$\,M$_\odot$). The resulting accretion rates are similar to matter accretion rates estimated in earlier works \cite[e.g.,][]{dekel09}.

As we are not primarily interested in the accretion of dark matter, but of the baryonic matter, we multiply the total matter accretion rate with the constant baryonic matter fraction to obtain the baryonic accretion rate onto halos $\dot{M}_{\rm baryon}(M_{\rm vir},z)$. This assumes a perfect mixing between dark and baryonic matter in the IGM. The resulting baryonic accretion rates are shown as solid lines in Fig.~\ref{fig:theory}.

It is interesting to note that these accretion curves show a similar shape as our derived net infall/inflow rates ($\widetilde{\psi}_{\rm HI \rightarrow H2}$, $\widetilde{\psi}_{\rm HII \rightarrow HI}$): the accretion rates rise from high redshift to about $z\,\sim\,2$ (depending on the virial masses considered). This increase in accretion to its peak value is dominated by gravitationally driven growth of the halo mass function. The subsequent decline towards $z\,=\,0$ is due to the fact that the Universe expands and to the gradual decrease in the availability of accretable (dark) matter\footnote{As pointed out by \cite{salcido18}, a Universe without an accelerated ($\Lambda$--dominated) expansion does not significantly change the accretion rates, i.e. the {\em accelerated} expansion of the Universe is not the reason for the observed decline in the accretion rates.}.

\subsubsection{Accretion onto central disks}
\label{accretion}
So far we have only considered the accretion of matter on a dark matter $halo$. We now compare these rates to the actual accretion to the central $disk$, and add our \ion{H}{2} net infall and \ion{H}{1} inflow rates to Fig.~\ref{fig:theory} (same curves as in Fig.~\ref{fig:flows}, but on a logarithmic scale). A comparison to the total baryonic matter that is being accreted onto galaxies (solid blue lines) immediately implies that the material that is needed for the infall/inflow rates can be easily accounted for: of the total matter that is being accreted onto the dark matter halos of galaxies, only about 10--30\% is needed to explain the infall/inflow rates that are inferred by the observations (Sec.~\ref{rates}). Consequently, the majority of the accreted baryons will not make it to the central galaxy `disks'. 

An extensive literature has addressed the question of how the material that is accreted onto dark matter halos ends up in the centers of galaxies. In the standard picture, baryons from the IGM accrete onto dark matter halos, converting their gravitational energy into kinetic energy, which is subsequently shock--heated to the virial temperature of the halo. In addition to the formation of this hot halo, a large body of work suggests that dense filaments permeate the halos, leading to the formation of cold streams that feed the cold gas reservoir, and thus star formation, in the centers of galaxies. This process, referred to as `cold mode accretion', occurs on timescales of order a free--fall or the dynamical crossing time of a spherical halo ($\sim$10$^8$\,yr, depending on mass), where the separation between the `cold' and `hot' phase is around $10^{5.5}$\,K. The fraction of the gas that is accreted in this cold phase depends on both the halo mass and the redshift, but it is the cold mode that appears to be the dominant accretion mechanism throughout all redshifts in most simulations for all but the most massive halos  \citep{keres05, dekel06, dekel09, pan19, vandevoort11, nelson13}. Once the gas is in a cool phase, it cools quickly to lower temperatures that are typical of the atomic/molecular interstellar medium, on timescales $<$10$^7$\,yr \citep[e.g., ][]{cornuault18}. Our observations do not allow to distinguish between the different accretion mechanisms (`cold' vs `hot').

We note that the decline (from the peak to $z\,=\,0$) in the baryonic accretion rate onto the halo (factor of a few) is smaller than that of our observationally derived net infall/inflow rates onto the disks (decline by almost an order of magnitude). This implies that additional mechanisms are suppressing the accretion of material, and these mechanisms become more dominant at $z\,<\,2$. In a simple picture, the baryonic material that does not make it to the galaxy centers is heated by a number of processes, e.g., shocks, photo--ionization, through, e.g. stellar and AGN feedback. This hot material has very long cooling times. Assuming a typical temperature $\sim 10^6$\,K and a density $\sim 10^{-5}$\,cm$^{-3}$, with substantial variation \citep[e.g.,][]{shull12, shull17, nicastro18}, the nominal bremsstrahlung cooling time is about $8.5\times 10^{11}$ yr, or more than sixty times the Hubble time \citep{rosati02}. 

We note that the cosmic density of AGN and star formation has decreased by about an order of magnitude since its peak, and will continue to do so in the future. Hence, feedback must play a less important role at late cosmic times. To explain the continued decline in the cosmic star--formation rate at late times, we conjecture that only the densest gas in IGM filaments has been able to cool and stream into galaxy potential wells, and that these dense regions have been effectively `tapped-out' over the eons. In this picture, most of the gas in the IGM that was predestined to fall into galaxies has done so already, and what is left will diffuse away with cosmic expansion\footnote{The situation is similar to the conclusion of the pioneering work by \citet{toomre72} on the pre--destiny of galaxy mergers, in which they conclude: `Hence one must presume that the partners in most cases were already bound to each other prior to their latest encounter.'}.

\section{Concluding remarks}

We have used measurements of the cosmic molecular gas density to put new constraints on the baryon cycle and the gas accretion process for gas that is gravitationally bound to galaxies. We find that the cosmic H$_2$ density is less than or equal to the cosmic \ion{H}{1} density over all times, briefly approaching equality at $z\,\sim\,1.5$. Below a redshift of $z\,\sim\,1.5$, the stellar mass density begins to dominate over all gas components (H$_2$ and \ion{H}{1}), and completely dominates the baryon content within the main body of galaxies by $z\,=\,0$. The average cosmic gas depletion time, defined as the molecular gas density divided by the star--formation rate density, is approximately constant above redshifts $z\,\gtrsim\,2$, and then increases slightly from $\tau_{\rm depl}\,\sim\,(4 \pm 2)\,\times\,10^{8}$\,yr at $z\,\sim\,2$ to $\tau_{\rm depl}\,=\,(7 \pm 3)\,\times\,10^{8}$\,yr at $z\,=\,0$. Significant accretion of gas onto galaxies is needed to form the bulk of the stellar mass at z\,$<$\,1.5: Assuming that the maximum molecular gas density (seen at $z\,\sim\,1.5$) {\rm will be fully transformed to stellar mass} can only account for at most a quarter of the stellar mass seen at $z\,=\,0$. 

The new H$_2$ constraints can be used to break up the gas accretion process onto galaxies in two steps. (i) First is the net inflow of atomic gas, and conversion to molecular gas, from the extended reservoirs to the centers of the dark matter halos (Equation~\ref{eq_rhodotinf}). (ii) Second is the net infall of mostly diffuse (ionized) gas to refuel the \ion{H}{1} reservoirs (Equation~\ref{eq_rhodotacc}). We find that both flow processes decrease sharply at redshifts  $z\,\lesssim\,1.5$, following, to first order, the star--formation rate density.

Zooming out, we can describe the gas cycle in galaxies as follows: an extended reservoir of atomic gas (\ion{H}{1}) is  formed by a (net) infall of  gas from the IGM/CGM at a rate of $\widetilde{\psi}_{\rm HII \rightarrow HI}$. This extended \ion{H}{1} component is not immediately associated with the star--formation process. Further (net) inflow from the \ion{H}{1}\ reservoir at a rate of $\widetilde{\psi}_{\rm HI \rightarrow H2}$ then leads to a molecular gas phase in the centers of the dark matter potentials. As the extended \ion{H}{1} density remains approximately constant, these two net rates are similar. Stars are then formed out of the molecular gas phase, and the resulting star--formation rate surface density in a galaxy is expected to be correlated with the molecular gas surface density \citep[e.g., ][]{bigiel08, leroy13}. The functional shape of the star--formation rate density $\psi_{\rm stars}$ is thus mostly driven by the availability of molecular gas, which in turn is defined by (net) infall rates of gas from larger distances. A comparison to numerical simulations shows that there is ample material that is being accreted onto dark matter halos. The decrease in gas accretion since $z\,\sim\,1.5$ is then a result of the decreased growth of dark matter halos (partly due to the expansion of the Universe), combined with the effects of feedback from stars and accreting black holes.

Lastly, by extrapolating our empirical fitting functions for the evolution of the stellar mass, \ion{H}{1}, and H$_2$, we find that the molecular gas density will decrease by about a factor of two in the next 5\,Gyr, the \ion{H}{1} mass density will remain approximately constant, and the stellar mass density will increase by about 10\%. The inflow and accretion rates will decrease correspondingly, and the cosmic star formation rate density will continue its steady descent to the infinitesimal.

\acknowledgments
 
We thank the referee for a very constructive report that helped to improve the paper. We thank Annalisa Pillepich and Andrea Ferrara for useful discussions. FW and MN acknowledge support from the ERC Advanced Grant 740246 (Cosmic\_Gas). BM acknowledges support from the Collaborative Research Centre 956, sub-project A1, funded by the Deutsche Forschungsgemeinschaft (DFG) -- project ID 184018867. TD-S acknowledges support from the CASSACA and CONICYT fund CAS-CONICYT Call 2018. RJA was supported by FONDECYT grant 1191124. DR acknowledges support from the National Science Foundation under grant numbers AST-1614213 and AST- 1910107 and from the Alexander von Humboldt Foundation through a Humboldt Research Fellowship for Experienced Researchers. IRS acknowledges support from STFC (ST/T000244/1). HI acknowledges support from JSPS KAKENHI Grant Number JP19K23462. JH acknowledges support of the VIDI research programme with project number 639.042.611, which is (partly) financed by the Netherlands Organisation for Scientific Research (NWO). DO is a recipient of an Australian Research Council Future Fellowship (FT190100083) funded by the Australian Government. This paper makes use of the following ALMA data: ADS/JAO.ALMA\# 2017.1.00118.S, ADS/JAO.ALMA\# 2015.1.01115.S. ALMA is a partnership of ESO (representing its member states), NSF (USA) and NINS (Japan), together with NRC (Canada), NSC and ASIAA (Taiwan), and KASI (Republic of Korea), in cooperation with the Republic of Chile. The Joint ALMA Observatory is operated by ESO, AUI/NRAO and NAOJ. The National Radio Astronomy Observatory is a facility of the National Science Foundation operated under cooperative agreement by Associated Universities, Inc.

\facility{ALMA}

\newpage

\appendix

\section{Background for Equations 3 and 4.}
We here provide some more background for the derivations of Equations~\ref{eq_rhodot_h2} and~\ref{eq_rhodothi} in the main body of the text. We consider the four main baryonic phases that are introduced in the schematic (Fig.~\ref{fig:schematic}). These are measured by their respective space densities: $\rho_{\rm stars}$ of  the gas in disks and bulges, $\rho_{\rm HII}$ of the ionized gas in the CGM and IGM, $\rho_{\rm HI}$ of the atomic gas within  galaxy disks and their environment, and $\rho_{\rm H2}$ of the molecular gas in galaxy disks. We neglect all other minor mass components discussed in Sec.~\ref{redshift0}. We can than express the phase evolution of the Universe in terms of 12 flow rates $\psi_{\rm x \rightarrow y}(z)$ (with $x,y$\,=\,[stars, HII, HI, H2], and $x\neq y$) that describe the following four phases $\rho_{\rm x}$. These flow rates are a function of redshift $z$, but for simplicity we omit the $(z)$ notation in the following.

\begin{align}
\dot{\rho}_{\rm stars} & = \underbrace{\psi_{\rm H2\rightarrow stars}  - \psi_{\rm stars \rightarrow H2}}_{\rm H2 \, \leftrightarrow \, stars} \underbrace{- \psi_{\rm stars \rightarrow HII}  + \psi_{\rm HII \rightarrow stars}}_{\rm HII\, \leftrightarrow \, stars}  \underbrace{- \psi_{\rm stars \rightarrow HI}  + \psi_{\rm HI \rightarrow stars}}_{\rm HI \, \leftrightarrow \, stars} \label{eq:rho_dot_star}\\[2mm]
\dot{\rho}_{\rm HII} & = \underbrace{\psi_{\rm H2\rightarrow HII}  - \psi_{\rm HII \rightarrow H2}}_{\rm H2 \, \leftrightarrow \, HII}  \underbrace{- \psi_{\rm HII \rightarrow HI}  + \psi_{\rm HI \rightarrow HII}}_{\rm HI \, \leftrightarrow \, HII}  \underbrace{- \psi_{\rm HII \rightarrow stars}  + \psi_{\rm stars \rightarrow HII}}_{\rm stars \, \leftrightarrow \, HII} \label{eq:rho_dot_hii}\\[2mm]
\dot{\rho}_{\rm HI} & = \underbrace{\psi_{\rm H2 \rightarrow HI}  - \psi_{\rm HI \rightarrow H2}}_{\rm H2 \, \leftrightarrow \, HI}  \underbrace{+ \psi_{\rm HII \rightarrow HI}  - \psi_{\rm HI \rightarrow HII}}_{\rm HII \, \leftrightarrow \, HI}   \underbrace{+ \psi_{\rm stars \rightarrow HI}  - \psi_{\rm HI \rightarrow stars}}_{\rm stars \, \leftrightarrow \, HI} \label{eq:rho_dot_hi}\\[2mm]
\dot{\rho}_{\rm H2} & = \underbrace{\psi_{\rm HII \rightarrow H2}  - \psi_{\rm H2 \rightarrow HII}}_{\rm HII \, \leftrightarrow \, H2} \underbrace{+ \psi_{\rm stars \rightarrow H2}  - \psi_{\rm H2 \rightarrow stars}}_{\rm stars \, \leftrightarrow \, H2}   \underbrace{+ \psi_{\rm HI \rightarrow H2}  - \psi_{\rm H2 \rightarrow HI}}_{\rm HI \, \leftrightarrow \, H2}\label{eq:rho_dot_h2}
\end{align}

\begin{figure*}[b]
\centering 
\includegraphics[width=0.35\textwidth, angle=-90]{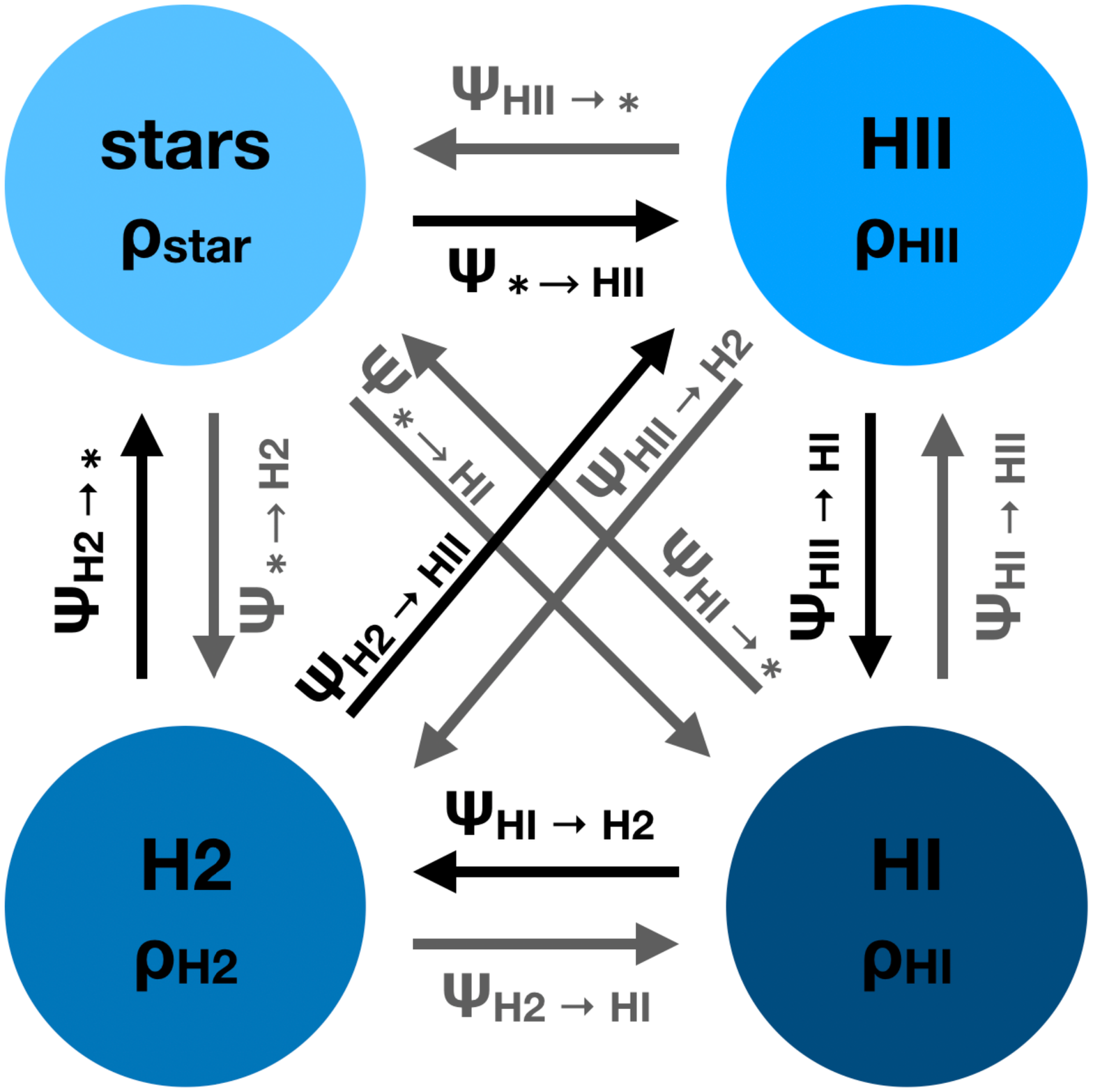}
\includegraphics[width=0.35\textwidth, angle=-90]{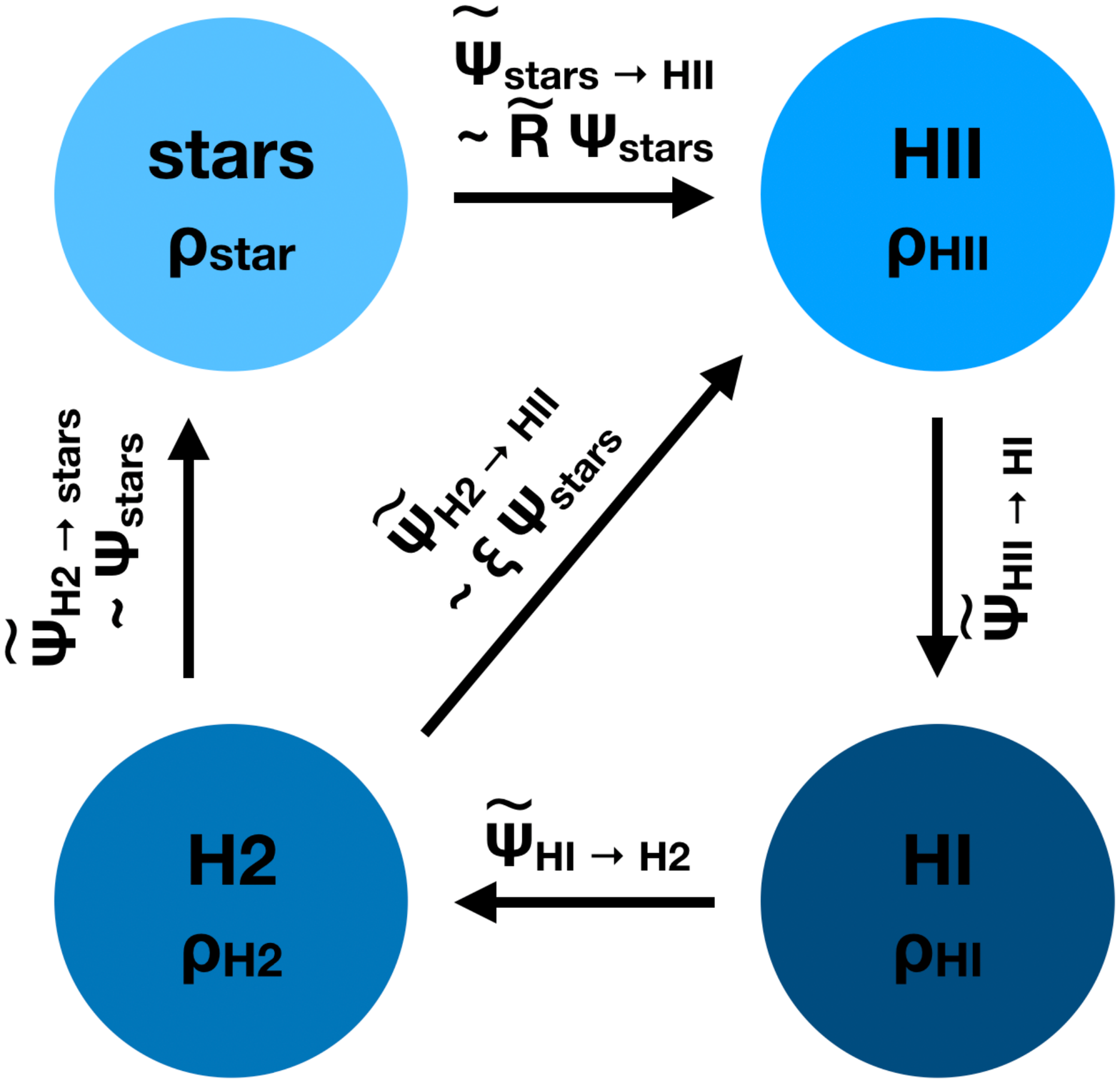}
\caption{Diagram of the flows between the different main baryonic phases (stars, ionized gas \ion{H}{2}, atomic gas \ion{H}{1}, molecular gas H$_2$) in the Universe. The phase evolution can be expressed with 12 flow rates $\psi_{\rm x \rightarrow y}$ that describe the 4 phases $\rho_{\rm x}$ as indicated in the left hand panel (equations \ref{eq:rho_dot_star}--\ref{eq:rho_dot_h2}). Darker arrows indicate the main flows in the diagram. The panel to the right shows how the flows are simplified, by setting some flow terms to zero (see text) and introducing {\em net flows} that are marked with a tilde.}
\label{fig:cycle1}
\end{figure*}

These flow rates are also visualized in Fig.~\ref{fig:cycle1}, and the `+' and `-' signs in the equations above denote `gains' and `losses', as indicated by the arrows in that figure.  We have measurements for three of these quantities, i.e. $\dot{\rho}_{\rm stars}$, $\dot{\rho}_{\rm HI}$, and $\dot{\rho}_{\rm H2}$, but many more unknowns, leaving the problem unsolvable. We can however simplify the above equations given our assumptions discussed in Sec.~\ref{schematic} and the corresponding schematic (Fig.~\ref{fig:schematic}). 

$\dot{\rho}_{\rm stars}$: In Eq.~\ref{eq:rho_dot_star}  the first two terms correspond to the mass flows between the stars and the H$_2$, the second two terms to the flows between the stars and the \ion{H}{2}, and the last two terms the flows between the stars and the \ion{H}{1}. We here assume that stars do not directly form out of \ion{H}{1} or \ion{H}{2}, and thus set these terms ($\psi_{\rm HI \rightarrow stars}$, $\psi_{\rm HII \rightarrow stars}$) to zero. Likewise, we assume that stars do not produce atomic hydrogen nor molecular gas, and we thus set these two terms ($\psi_{\rm stars \rightarrow HI}$, $\psi_{\rm stars \rightarrow H2}$) to zero as well. Equation.~\ref{eq:rho_dot_star} therefore simplifies to:

\begin{equation}
\dot{\rho}_{\rm stars}  = \psi_{\rm H2\rightarrow stars} - \psi_{\rm stars \rightarrow HII} \label{eq:rho_dot_star_simp}
\end{equation}

$\dot{\rho}_{\rm HII}$: In Eq.~\ref{eq:rho_dot_hii}  the first two terms correspond to the mass flows between the H$_2$ and the ionized gas (\ion{H}{2}), the second two terms to the flows between the atomic and ionized gas, and the last two terms the flows between the stars and the ionized gas. We here assume that ionized gas cannot directly form molecular gas, and that ionized gas can not directly form stars and set $\psi_{\rm HII \rightarrow stars}$ and $\psi_{\rm HII \rightarrow H2}$ to zero. The other flows are in principle plausible, and equation \ref{eq:rho_dot_hii} thus becomes:

\begin{align}
\dot{\rho}_{\rm HII} & = \psi_{\rm H2\rightarrow HII}  - \psi_{\rm HII \rightarrow HI}  + \psi_{\rm HI \rightarrow HII}  + \psi_{\rm stars \rightarrow HII} \label{eq:rho_dot_hii_simp}\\
\dot{\rho}_{\rm HII} & = \psi_{\rm H2\rightarrow HII}  - \widetilde{\psi}_{\rm HII \rightarrow HI} + \psi_{\rm stars \rightarrow HII}   \label{eq:rho_dot_hii_simp2}
\end{align}
where in a second step we introduced $\widetilde{\psi}_{\rm HII \rightarrow HI}$ as being the {\em net} flow between the ionized gas phase and the atomic gas. As we assume that $|\psi_{\rm HII \rightarrow HI}| > |\psi_{\rm HI \rightarrow HII}|$ we assign a  minus sign to the net flow $\widetilde{\psi}_{\rm HII \rightarrow HI}$ in Eq.~\ref{eq:rho_dot_hii_simp2}. 

$\dot{\rho}_{\rm HI}$: In Eq.~\ref{eq:rho_dot_hi}  the first two terms correspond to the mass flows between the H$_2$ and the atomic gas, the second two terms to the flows between the atomic and ionized gas, and the last two terms to the flows between the stars and the atomic gas. As before, we assume that no stars can form out of \ion{H}{1}, and that stars cannot form observable (i.e. long--lived) \ion{H}{1}, i.e. both $\psi_{\rm HI \rightarrow stars}$ and $\psi_{\rm stars \rightarrow HI}$ are set to zero. Eq.~\ref{eq:rho_dot_hi} then yields:

\begin{align}
\dot{\rho}_{\rm HI} & = \psi_{\rm H2 \rightarrow HI}  - \psi_{\rm HI \rightarrow H2}  + \psi_{\rm HII \rightarrow HI}  - \psi_{\rm HI \rightarrow HII}  \label{eq:rho_dot_hi_simp}\\
\dot{\rho}_{\rm HI} & =  - \widetilde{\psi}_{\rm HI \rightarrow H2}  + \widetilde{\psi}_{\rm HII \rightarrow HI}  \label{eq:rho_dot_hi_simp2}
\end{align}
where in a second step we again introduce an additional {\em net} flow between the atomic gas and the H$_2$ ($\widetilde{\psi}_{\rm HI \rightarrow H2}$). Its sign is determined as we assume $|\psi_{\rm H2 \rightarrow HI}| < |\psi_{\rm HI \rightarrow H2}|$. This equation is identical to Eq.~\ref{eq_rhodothi} in the main body of this manuscript.

$\dot{\rho}_{\rm H2}$: In Eq.~\ref{eq:rho_dot_h2}  the first two terms correspond to the mass flows between the H$_2$ and the ionized gas, the second two terms to the flows between the H$_2$ and the stars, and the last two terms to the flows between the H$_2$ and the atomic gas. As before, we set $\psi_{\rm HII \rightarrow H2}$ and $\psi_{\rm stars \rightarrow H2}$ to zero. This yields:

\begin{align}
\dot{\rho}_{H2} & =  - \psi_{\rm H2 \rightarrow HII}    - \psi_{\rm H2 \rightarrow stars}  + \psi_{\rm HI \rightarrow H2}  - \psi_{\rm H2 \rightarrow HI}\label{eq:rho_dot_h2_simp} \\
\dot{\rho}_{H2} & =  - \psi_{\rm H2 \rightarrow HII}    - \psi_{\rm H2 \rightarrow stars}  + \widetilde{\psi}_{\rm HI \rightarrow H2} \label{eq:rho_dot_h2_simp2} 
\end{align}
where we again use the previously defined net flow between the atomic gas and the H$_2$, $\widetilde{\psi}_{\rm HI \rightarrow H2}$, in the second step. 

We can now further simplify equation ~\ref{eq:rho_dot_h2_simp2} by setting $\psi_{\rm H2 \rightarrow stars}$ to the star formation rate density, i.e. $\psi_{\rm H2 \rightarrow stars} = \psi_{\rm stars}$, i.e. the fundamental process that forms stars out of molecular gas. We can also set the feedback rate (molecular gas that will be ionized), $\psi_{\rm H2 \rightarrow HII}$, to be proportional to the star formation rate density, i.e. $\psi_{\rm H2 \rightarrow HII} = \xi \psi_{\rm stars}$. This yields Eq.~\ref{eq_rhodot_h2} in the main text.

As a side note, assuming that the return rate of the stars to the ionized medium ($\psi_{\rm stars \rightarrow HII}$) is proportional to the star formation rate density $\psi_{\rm stars}$ with a proportionality factor $\tilde{R}$, i.e. $\psi_{\rm stars \rightarrow HII} = \tilde{R} \psi_{\rm stars}$, we can rewrite Eq.~\ref{eq:rho_dot_star_simp} that expresses the change in the stellar mass density as follows:

\begin{align}
\dot{\rho}_{\rm stars} & = \psi_{\rm stars} - \tilde{R} \psi_{\rm stars} \\
                       & = (1 - \tilde{R}) \psi_{\rm stars}
\label{eq:rho_dot_star_simp2}
\end{align}

Note that the factor $\tilde{R}$ would be equal to the classical return factor $R$ (see Sec.~\ref{stars}) if we assume that all loss of stellar mass (stellar winds, SN explosions) would end up in the ionized phase of the interstellar medium.

\newpage
\section{Observational data for the cosmic density measurements.}
In this appendix, we give the observational measurements for both the cosmic \ion{H}{1} mass density and the  cosmic H$_2$ mass density from the literature. The observational data used to fit the cosmic stellar mass density and the cosmic star formation rate density are taken from the compilation in \citet{madau14}.

\begin{deluxetable*}{cccc}[b]
\tablecaption{Measurements of the cosmic H$_2$ mass density
\label{tab:h2obs}}
\tablehead{
\colhead{Redshift} & 
\colhead{$\rho_{\rm H2}$} &
\colhead{Method} &
\colhead{Reference} \\
\colhead{} &
\colhead{($10^{8} M_\odot$ Mpc$^{-3}$)} &
\colhead{} &
\colhead{}
} 
\startdata
$0.01 - 0.05$ & $0.104 _{-0.009}^{+0.009}$ & CO/xCOLD GASS & \citet{fletscher20} \\
$0.00 - 0.37^a$ & $0.009 _{-0.007}^{+0.019}$ & CO/ASPECS & \citet{decarli19} \\
$0.27 - 0.63$ & $0.11 _{-0.05}^{+0.10}$ & CO/ASPECS & \citet{decarli20} \\
$0.69 - 1.17$ & $0.46 _{-0.18}^{+0.27}$ & CO/ASPECS & \citet{decarli20} \\
$1.01 - 1.74$ & $0.55 _{-0.15}^{+0.20}$ & CO/ASPECS & \citet{decarli19} \\
$2.01 - 3.11$ & $0.29 _{-0.11}^{+0.15}$ & CO/ASPECS & \citet{decarli19} \\
$3.01 - 4.47$ & $0.24 _{-0.07}^{+0.09}$ & CO/ASPECS & \citet{decarli19} \\
$0.48 - 1.48$ & $0.41 _{-0.12}^{+0.11}$ & CO/PHIBBS2 & \citet{lenkic20} \\
$1.01 - 2.01$ & $0.48 _{-0.11}^{+0.12}$ & CO/PHIBBS2 & \citet{lenkic20} \\
$2.01 - 3.01$ & $0.37 _{-0.11}^{+0.10}$ & CO/PHIBBS2 & \citet{lenkic20} \\
$3.14 - 4.14$ & $0.15 _{-0.07}^{+0.07}$ & CO/PHIBBS2 & \citet{lenkic20} \\
$4.25 - 5.25^b$ & $0.10 _{-0.06}^{+0.07}$ & CO/PHIBBS2 & \citet{lenkic20} \\
$1.95 - 2.85$ & $0.27 _{-0.11}^{+0.16}$ & CO/COLDz & \citet{riechers19} \\
$4.9 - 6.7^b$ & $0.047 _{-0.023}^{+0.034}$ & CO/COLDz & \citet{riechers19} \\
$1.95 - 2.85$ & $0.28 _{-0.12}^{+0.18}$ & CO/VLASPECS & \citet{riechers20} \\
\hline
$0.3 - 0.6$ & $0.20 _{-0.14}^{+0.16}$ & Dust continuum & \citet{magnelli20} \\
$0.6 - 1.0$ & $0.30 \pm 0.18$ & Dust continuum & \citet{magnelli20} \\
$1.0 - 1.6$ & $0.51 \pm 0.11$ & Dust continuum & \citet{magnelli20} \\
$1.6 - 2.3$ & $0.33 \pm 0.07$ & Dust continuum & \citet{magnelli20} \\
$2.3 - 3.2$ & $0.42 \pm 0.09$ & Dust continuum & \citet{magnelli20} \\
$0.2 - 0.5$ & $0.43 _{-0.05}^{+0.06}$ & Dust continuum & \citet{scoville17} \\
$0.5 - 0.8$ & $0.47 _{-0.07}^{+0.08}$ & Dust continuum & \citet{scoville17} \\
$0.8 - 1.1$ & $0.59 _{-0.09}^{+0.10}$ & Dust continuum & \citet{scoville17} \\
$1.1 - 1.5$ & $0.84 _{-0.13}^{+0.14}$ & Dust continuum & \citet{scoville17} \\
$1.5 - 2.0$ & $0.90 _{-0.13}^{+0.16}$ & Dust continuum & \citet{scoville17} \\
$2.0 - 2.5$ & $0.74 _{-0.11}^{+0.12}$ & Dust continuum & \citet{scoville17} \\
$2.5 - 3.0$ & $0.63 _{-0.09}^{+0.11}$ & Dust continuum & \citet{scoville17} \\
$3.0 - 4.0$ & $0.36 _{-0.05}^{+0.06}$ & Dust continuum & \citet{scoville17} \\
$0.2 - 0.6$ & $0.23 _{-0.05}^{+0.06}$ & Dust continuum & \citet{berta13} \\
$0.7 - 1.0$ & $0.60 _{-0.17}^{+0.18}$ & Dust continuum & \citet{berta13} \\
$1.0 - 2.0$ & $0.48 _{-0.07}^{+0.09}$ & Dust continuum & \citet{berta13} \\
\enddata
\tablenotetext{a}{The lowest redshift bin from the ASPECS survey is not used in the fit due to the small cosmic volume probed \citep{decarli20}.}
\tablenotetext{b}{As we only consider redshifts $z\,<\,4$ in this study, these measurements are not used in our fit.}
\tablecomments{The table has been subdivided into categories based on the observational method used to estimate $\rho_{\rm H2}$, either through measurement of a CO line (from a blind survey or from targeted observations of individual galaxies), or through the measurement of the dust continuum.}
\end{deluxetable*}

\begin{deluxetable*}{cccc}
\tablecaption{Measurements of the cosmic \ion{H}{1} mass density
\label{tab:hiobs}}
\tablehead{
\colhead{Redshift} & 
\colhead{$\rho_{\rm HI}$} &
\colhead{Method} &
\colhead{Reference} \\
\colhead{} &
\colhead{($10^{8} M_\odot$ Mpc$^{-3}$)} &
\colhead{} &
\colhead{}
} 
\startdata
$0.0$ & $0.60 \pm 0.10$ & 21\,cm & \citet{zwaan05} \\
$0.0$ & $0.51 \pm 0.09$ & 21\,cm & \citet{jones18} \\
$0.0$ & $1.03 \pm 0.17$ & 21\,cm & \citet{braun12} \\
\hline
$0.028$ & $0.65 _{-0.06}^{+0.13}$ & 21\,cm-stacked & \citet{delhaize13} \\
$0.096$ & $0.73 _{-0.10}^{+0.13}$ & 21\,cm-stacked & \citet{delhaize13} \\
$0.06$ & $0.37 \pm 0.11$ & 21\,cm-stacked & \citet{hoppmann15} \\
$0.1$ & $0.53 \pm 0.08$ & 21\,cm-stacked & \citet{rhee13} \\
$0.2$ & $0.55 \pm 0.14$ & 21\,cm-stacked & \citet{rhee13} \\
$0.24$ & $1.11 \pm 0.49$ & 21\,cm-stacked & \citet{lah07} \\
$0.2 - 0.4$ & $0.62 \pm 0.09$ & 21\,cm-stacked &\citet{bera19} \\
$1.265$ & $< 0.342$ & 21\,cm-stacked & \citet{kanekar16} \\
\hline
$0.2 - 1.8$ & $0.99 _{-0.36}^{+0.55}$ & 21\,cm cross-correlation &\citet{masui13} \\
\hline
$0.11 - 0.61$ & $1.11 \pm 0.37$ & \ion{Mg}{2}-selection & \citet{rao17} \\
$0.61 - 0.89$ & $0.96 \pm 0.23$ & \ion{Mg}{2}-selection & \citet{rao17} \\
$0.89 - 1.65$ & $1.08 \pm 0.42$ & \ion{Mg}{2}-selection & \citet{rao17} \\
\hline
$0.01 - 0.48$ & $0.40 _{-0.19}^{+0.31}$ & DLA & \citet{shull17} \\
$0.01 - 1.6$ & $0.33 _{-0.16}^{+0.26}$ & DLA & \citet{neeleman16} \\
$1.55 - 2.0$ & $1.01 _{-0.29}^{+0.34}$ & DLA & \citet{peroux03} \\
$2.0 - 2.7$ & $1.45 _{-0.30}^{+0.34}$ & DLA & \citet{peroux03} \\
$2.7 - 3.5$ & $1.33 _{-0.30}^{+0.35}$ & DLA & \citet{peroux03} \\
$3.5 - 4.85$ & $0.82 _{-0.27}^{+0.30}$ & DLA & \citet{peroux03} \\
$1.55 - 2.73$ & $2.13 _{-0.86}^{+1.28}$ & DLA & \citet{sanchez-ramirez16} \\
$2.73 - 3.21$ & $0.79 _{-0.38}^{+0.63}$ & DLA & \citet{sanchez-ramirez16} \\
$3.21 - 4.5$ & $1.78 _{-0.42}^{+0.49}$ & DLA & \citet{sanchez-ramirez16} \\
$2.0 - 2.3$ & $1.42 \pm 0.07$ & DLA & \citet{noterdaeme12} \\
$2.3 - 2.6$ & $1.25 \pm 0.06$ & DLA & \citet{noterdaeme12} \\
$2.6 - 2.9$ & $1.50 \pm 0.07$ & DLA & \citet{noterdaeme12} \\
$2.9 - 3.2$ & $1.58 \pm 0.11$ & DLA & \citet{noterdaeme12} \\
$3.2 - 3.5$ & $1.83 \pm 0.19$ & DLA & \citet{noterdaeme12} \\
$2.2 - 2.4$ & $0.74 \pm 0.13$ & DLA & \citet{prochaska09} \\
$2.4 - 2.7$ & $1.00 \pm 0.11$ & DLA & \citet{prochaska09} \\
$2.7 - 3.0$ & $1.00 \pm 0.10$ & DLA & \citet{prochaska09} \\
$3.0 - 3.5$ & $1.40 \pm 0.12$ & DLA & \citet{prochaska09} \\
$3.5 - 4.0$ & $1.62 \pm 0.28$ & DLA & \citet{prochaska09} \\
$4.0 - 5.5$ & $1.58 _{-0.30}^{+0.34}$ & DLA & \citet{prochaska09} \\
$2.55 - 3.4$ & $1.48 _{-0.50}^{+0.66}$ & DLA & \citet{guimaraes09} \\
$3.4 - 3.83$ & $1.31 _{-0.45}^{+0.51}$ & DLA & \citet{guimaraes09} \\
$3.83 - 5.03$ & $1.16 _{-0.42}^{+0.52}$ & DLA & \citet{guimaraes09} \\
$3.56 - 4.45$ & $1.87 _{-0.40}^{+0.42}$ & DLA & \citet{crighton15} \\
$4.45 - 5.31$ & $1.56 _{-0.27}^{+0.31}$ & DLA & \citet{crighton15}
\enddata
\tablecomments{Values have been converted, where necessary, to the adopted cosmology and have been corrected for the contribution of helium. For the DLA measurements, no correction for lower \ion{H}{1} column density systems has been applied. The table has been subdivided into categories based on the observational method used to estimate $\rho_{\rm HI}$.}
\end{deluxetable*}

\end{document}